\documentclass[%
 reprint,
superscriptaddress,
 amsmath,amssymb,
 aps,
prb,citeautoscript
]{revtex4-2}

\usepackage{graphicx}
\usepackage{dcolumn}
\usepackage{bm}
\usepackage{ragged2e}
\usepackage[utf8]{inputenc}
\usepackage[T1]{fontenc}
\usepackage{mathptmx}
\usepackage{etoolbox}
\usepackage{svg}
\usepackage{graphics,color,epsf,float,bm}
\usepackage{siunitx}

\begin{document}

\title{Exploring van der Waals cuprate superconductors using a hybrid microwave circuit}

\author{Haolin Jin}
\thanks{These authors contributed equally to this work.}
\affiliation{Max Planck Institute for Chemical Physics of Solids, 01187 Dresden, Germany}
\affiliation{Institute of Solid State and Material Physics, Technische Universit{\"a}t Dresden, 01062 Dresden, Germany}

\noindent

\author{Giuseppe Serpico}
\thanks{These authors contributed equally to this work.}
\affiliation{Max Planck Institute for Chemical Physics of Solids, 01187 Dresden, Germany}
\affiliation{Department of Physics, University of Naples Federico II, Via Cintia, 80126 Naples, Italy}

\author{Yejin Lee}
\affiliation{Max Planck Institute for Chemical Physics of Solids, 01187 Dresden, Germany}

\author{Tommaso Confalone}
\affiliation{Leibniz Institute for Solid State and Materials Science Dresden (IFW Dresden), 01069 Dresden, Germany}
\affiliation{Institute of Applied Physics, Technische Universit{\"a}t Dresden, 01062 Dresden, Germany}

\author{Christian N. Saggau}
\affiliation{Leibniz Institute for Solid State and Materials Science Dresden (IFW Dresden), 01069 Dresden, Germany}
 \affiliation{DTU Electro, Department of Electrical and Photonics Engineering, Technical University of Denmark, 2800 Kongens Lyngby, Denmark}
\affiliation{Center for Silicon Photonics for Optical Communications (SPOC), Technical University of Denmark, 2800 Kongens Lyngby, Denmark}

\author{Flavia Lo Sardo}
\affiliation{Leibniz Institute for Solid State and Materials Science Dresden (IFW Dresden), 01069 Dresden, Germany}
\affiliation{Institute of Materials Science, Technische Universit{\"a}t Dresden, 01062 Dresden, Germany}

\author{Genda Gu}
\affiliation{Condensed Matter Physics and Materials Science Department, Brookhaven National Laboratory, Upton, NY 11973, USA}

\author{Berit H. Goodge}
\affiliation{Max Planck Institute for Chemical Physics of Solids, 01187 Dresden, Germany}

\author{Edouard Lesne}
\affiliation{Max Planck Institute for Chemical Physics of Solids, 01187 Dresden, Germany}

\author{Domenico Montemurro}
\affiliation{Department of Physics, University of Naples Federico II, Via Cintia, 80126 Naples, Italy}

\author{Kornelius Nielsch}
\affiliation{Leibniz Institute for Solid State and Materials Science Dresden (IFW Dresden), 01069 Dresden, Germany}
\affiliation{Institute of Applied Physics, Technische Universit{\"a}t Dresden, 01062 Dresden, Germany}
\affiliation{Institute of Materials Science, Technische Universit{\"a}t Dresden, 01062 Dresden, Germany}

\author{Nicola Poccia}
\affiliation{Department of Physics, University of Naples Federico II, Via Cintia, 80126 Naples, Italy}
\affiliation{Leibniz Institute for Solid State and Materials Science Dresden (IFW Dresden), 01069 Dresden, Germany}

\author{Uri Vool}
\thanks{uri.vool@cpfs.mpg.de}
\affiliation{Max Planck Institute for Chemical Physics of Solids, 01187 Dresden, Germany}
\affiliation{Leibniz Institute for Solid State and Materials Science Dresden (IFW Dresden), 01069 Dresden, Germany}

\makeatletter
\renewcommand{\fnum@figure}{\textbf{Figure~\thefigure}}
\makeatother

\begin{abstract}
\section*{abstract}

The advent of two-dimensional van der Waals materials is a frontier of condensed matter physics and quantum devices.
However, characterizing such materials remains challenging due to the limitations of bulk material techniques, necessitating the development of specialized methods.
Here, we investigate the superconducting properties of $\text{Bi}_{2}\text{Sr}_{2}\text{CaCu}_{2}\text{O}_{8+x}$ flakes by integrating them with a hybrid superconducting microwave resonator.  
The hybrid resonator is significantly modified by the interaction with the flake while maintaining a high quality factor ($3 \times 10^4$).
We also observe a significant upshift of the resonator frequency with increasing temperature, as well as a positive nonlinearity.
These effects originate from a presently unknown microscopic mechanism within the flake, and can be modeled as a two-level system bath interacting with the resonant mode.
Our findings open a path for high quality hybrid circuits with van der Waals flakes for exploring novel materials and developing new devices for quantum technology.
\noindent
\\
\end{abstract}

\maketitle

Two-dimensional van-der-Waals (vdW) materials are a rapidly growing field, with new materials exhibiting a wide range of phenomena, including ballistic transport \cite{du2008approaching,mayorov2011micrometer}, magnetism \cite{gong2017discovery,huang2018electrical,wang2022magnetic},  topology \cite{deng2020quantum,wu2018observation,fatemi2018electrically} and unconventional superconductivity \cite{xi2016ising,zhao2019sign,yu2019high,meng2024layer}. 
A key advantage of vdW materials is their versatility in tuning physical properties through in-situ electrical gating and the creation of novel material systems by stacking different material layers and varying their relative angle \cite{cao2018unconventional, kang2024evidence}.

However, a major challenge arises in the characterization and measurement of vdW flakes because many of the techniques available for exploring bulk materials cannot be directly applied to 2D flakes. 
Their reduced dimensionality, small size, and delicate nature pose significant challenges, necessitating the development and adaptation of specialized techniques.
For instance, the London penetration depth measurement is a common technique to uncover the superconducting gap symmetry \cite{hosseini1999microwave,prozorov2006magnetic}, but it requires large, pure crystal samples and is inapplicable for atomically thin and micrometer-sized flakes.

An alternative measurement technique has been recently developed to integrate the materials into a hybrid superconducting resonator \cite{thiemann2018single,Phan2022detecting,bottcher2023circuit}.  
Superconducting resonators are coherent macroscopic devices with high tunability and low-loss operation \cite{hammer2008ultra,gao2008physics,megrant2012planar,mahashabde2020fast},
making them particularly well-suited for quantum technology \cite{krantz2019quantum,blais2021circuit} as well as sensing applications. Indeed, recently superconducting resonators have been used in combination with vdW materials
to investigate their microwave losses \cite{antony2021making}, dielectric properties \cite{wang2022hexagonal,maji2024superconducting}, kinetic inductance \cite{kreidel2024measuring,tanaka2024superfluid}, and coupling to novel Josephson junctions\cite{schmidt2018ballistic, wang2019coherent, haller2022phase, butseraen2022gate}. 

\begin{figure*}[!t]
    \includegraphics{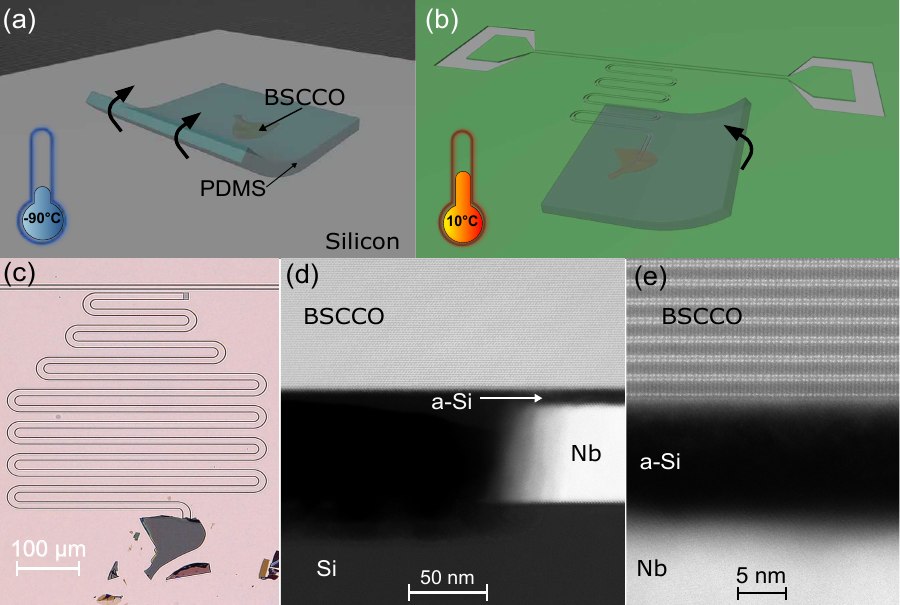}
    \caption{\textbf{Fabrication and characterization of a hybrid circuit}. (a) A BSCCO flake is picked up using a cold PDMS stamp at -90°C. (b) The flake is placed on top of the niobium resonator, and the temperature is raised to 10°C to remove the PDMS. (c) Optical image showing the completed hybrid device. (d) Scanning Transmission Electron Microscopy (STEM) image capturing the interface between BSCCO and the niobium at the edge of the central stripe of the resonator. (e) Close-up of the niobium-BSCCO cross-section showing atomically pristine BSCCO layers at the interface.}
    \label{fig1}      
\end{figure*}

But the fabrication of a hybrid device composed of a superconducting resonator and a vdW flake presents significant challenges. 
Superconducting resonators have been meticulously optimized over the years through careful selection of materials and fabrication processes to achieve high coherence.
Similarly, the techniques to isolate and preserve pristine vdW flakes have been carefully refined.
The interface between these devices can introduce imperfections and compromise both the coherence of the resonator and the structure of the flake.
For this reason, the fabrication of hybrid devices requires a dedicated controlled procedure and this is particularly true for complex and sensitive materials.
A particularly sensitive but remarkable vdw material is  $\text{Bi}_{2}\text{Sr}_{2}\text{CaCu}_{2}\text{O}_{8+x}$ (BSCCO).
BSCCO has garnered significant attention in recent years due to the experimental advances in preserving nearly perfect lattice and superconductivity in the atomically thin limit \cite{zhao2019sign,yu2019high} and for the realization of ultra-clean interfaces of twisted vdw heterostructures \cite{zhao2023time, lee2023encapsulating, martini2023twisted}.
This has led to a flourishing of theoretical predictions for new quantum states of matter \cite{can2021high,song2022doping,liu2023charge,yuan2023inhomogeneity} and further evidence for a dominant d-wave order parameter \cite{talantsev2024evidences}. 

In this work, we explore the microwave properties of an optimally doped vdW BSCCO flake by integrating it into a hybrid superconducting resonator.
We utilized a cryogenic stacking technique in a controlled environment, thus ensuring that the pristine structure of the crystal remains unharmed during sample preparation.
The resonant mode is strongly modified by the presence of the vdW flake, indicating a strong hybridization, while maintaining a high-quality factor ($3\times 10^{4}$). 
The hybrid circuit's microwave response is noticeably different from the bare resonator, showing a substantial increase of the resonance frequency with temperature. 
This behavior can be modeled as a two-level system (TLS) bath interacting with the resonator.
However, the resonator coherence which remains high despite the strong coupling to the TLS bath deviates from the model prediction. 
Furthermore, the hybrid circuit shows a strong positive nonlinearity, inconsistent with typical resonator nonlinearity due to kinetic inductance.
These effects are inherent within the BSCCO flake and possibly correspond to an off-resonant TLS bath, though their exact microscopic origin requires further investigation. 

\begin{figure*}[!t]
    \includegraphics[width=6.8in]{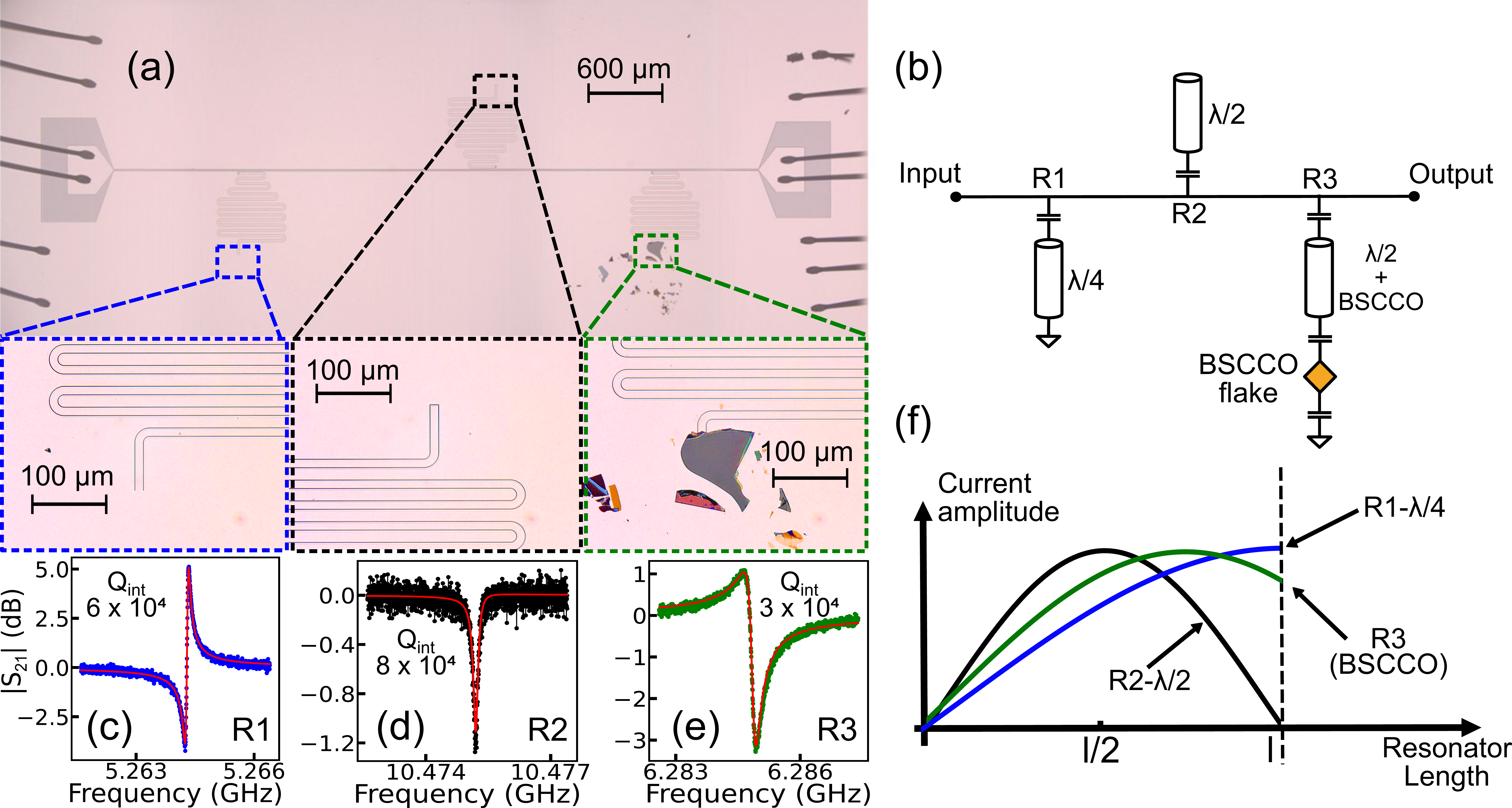}
    \caption{\textbf{Optical micrography and transmission spectrum}. (a)  Optical microscope image of a transmission line coupled to three co-planar resonators of equal length but with different boundary conditions: resonator R1 (conventional half-wave, left), R2 (conventional quarter-wave, middle) and R3 (half-wave shunted by a BSCCO flake, right), with enlarged images of the respective resonator terminations. (b) Diagram of the transmission line and resonators. (c-e) Low-power transmission spectrum of the resonator R1 (left, blue), R2 (middle, black) and R3 (with BSCCO flake, right, green) and their internal quality factors at 50 mK. (f) The distribution of current amplitude along the length of the resonator for the different resonant modes.}
    \label{fig2}     
\end{figure*}

Superconductivity in a BSCCO flake is crucially affected by changes of the spatial configurations of oxygen dopants, which become mobile above $\SI{-73}{\celsius}$ \cite{zeljkovic2012imaging,poccia2020spatially,figueruelo2024apparent}. 
Therefore, we employ a cryogenic transfer technique that preserves the spatially correlated superlattice order and freezes oxygen defects in their original positions \cite{lee2023encapsulating, martini2023twisted}.
We transferred the pre-exfoliated BSCCO flake onto the superconducting resonator using a polydimethylsiloxane (PDMS) polymer. 
PDMS exhibits increased adhesion at lower temperatures, making it suitable for cryogenic applications. 
The transfer process was conducted while the sample was cooled to -$\SI{90}{\celsius}$, freezing oxygen dopants (Figure \ref{fig1}a). 
The temperature of the sample stage was then raised to $\SI{10}{\celsius}$ (Figure \ref{fig1}b), allowing us to detach the PDMS while minimizing mechanical stress. The 450 nm flake was thus placed at the end of a co-planar resonator made out of 60 nm thick niobium thin film (Figure \ref{fig1}c).
To demonstrate the quality of the device at the atomic resolution, we use scanning transmission electron microscopy (STEM) on a cut where the BSCCO flake is held above the niobium in the gap between the center line and the ground plane (Figure \ref{fig1}d). The BSCCO flake maintains perfect crystalline structure with no degradation layer at the interface (Figure \ref{fig1}e). The pristine contact is improved by the amorphous silicon capping layer placed on the niobium, preventing its oxidation (See Supporting Information).

To quantify the effect of the BSCCO flake on the superconducting resonator, we designed a coplanar waveguide device composed of three different resonator types on the same chip (Figure \ref{fig2}a): a quarter-wave resonator shorted to the ground plane at the end (R1, $\lambda/4$), a half-wave resonator with open boundary conditions on both beginning and end (R2, $\lambda/2$), and a half-wave resonator which is shorted to ground by a BSCCO flake placed on it (R3), as illustrated in Figure \ref{fig2}b.
All three resonators are of 12 $\mathrm{\mu m}$ center line width, 1 $\mathrm{\mu m}$ gap width, and equal lengths, and are coupled to a single 50 $\Omega$ transmission line, ensuring consistency in sample quality and minimal lithographic process variations. 
The scattering parameter across the transmission line was measured while the sample was cooled to 50 mK, revealing resonance peaks corresponding to the three devices, as shown in Figure \ref{fig2}(c-e). 
The resonance frequency depends on the resonance wave length $\lambda$ as $f_{\text{res}} \propto 1/{\lambda}$. In this sense, the resonance frequency of the half-wave resonator mode ($\lambda/2= l$) is expected to be twice that of the quarter-wave resonator ($\lambda/4= l$), consistent with resonators R1 and R2. 
The resonator R3 with a BSCCO flake shorting the center line to ground resulted in a resonance frequency of 6.28 GHz, a substantial reduction from the half-wave resonance frequency of 10.47 GHz, showing the strong participation of the BSCCO flake in the resonant mode. 
This intermediate frequency suggests that the resonant mode with BSCCO lies between a $\lambda/4$ and $\lambda/2$ mode, as depicted in Figure \ref{fig2}f. This mode can be modeled assuming a finite capacitance between the BSCCO flake and the niobium resonator, which was numerically simulated to be approximately 5 pF (See Supporting Information). 
Concerning the quality factor, both R1 ($Q_{int} = 6 \times 10^{4}$) and R2 ($Q_{int} = 8 \times 10^{4}$) exhibit higher values compared to R3 ($Q_{int} = 3 \times 10^{4}$). The bound on the loss due to hybridization with the flake can be estimated as $Q_{flake}=(1/{{Q_{R3}}-1/{Q_{R2}}})^{-1} = 5\times 10^{4}$. 
The discrepancy between the quality factors could possibly be attributed to imperfections in the flake \cite{sahu2019nanoelectromechanical}, but the general high quality of the hybrid device indicates the BSCCO flake in our device maintains good superconducting properties. 
Additionally, we observed that silicon capping enhances the quality of the hybrid device compared to the resonators without silicon capping (See Supporting Information). 
This is possibly due to the mitigation of surface defects that degrade the coupling with the BSCCO \cite{ghosh2020demand}. 

\begin{figure}[h!]
\vspace{-0.2cm}
	\begin{center}
		\includegraphics[scale=0.3]{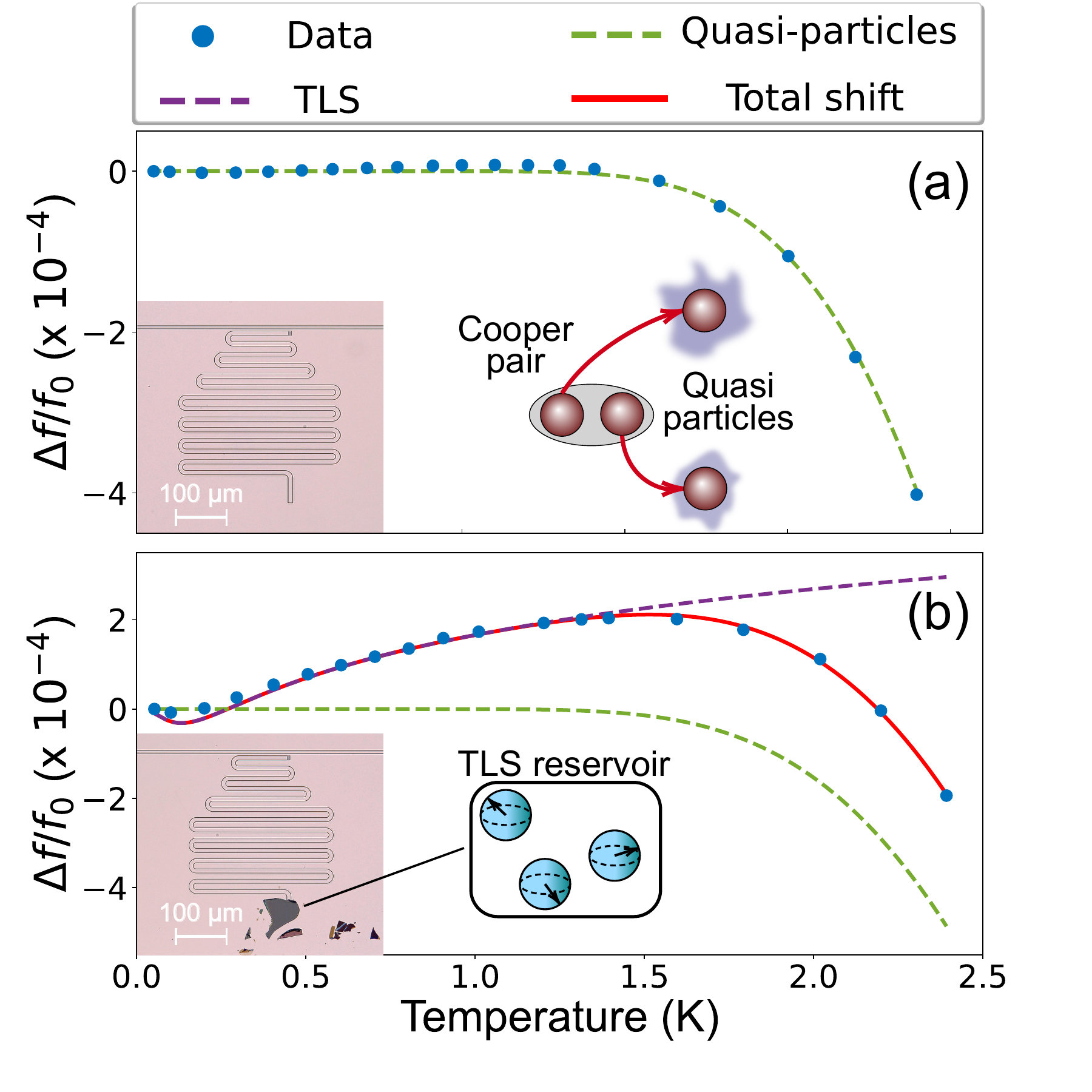}
        \caption{\textbf{Temperature dependence of the resonance frequency}. (a-b) Fractional change of the resonance frequency as a function of temperature for: Resonator R2, $\lambda/2$ niobium resonator (a) and resonator R3, $\lambda/2$ niobium resonator with a BSCCO flake (b). The green dashed line is a fit to a shift due to the thermal generation of quasiparticles, while the purple dashed line is a fit to coupling with a TLS reservoir. The red line shows a fit to a combined model of both.}
		\label{fig3}          
	\end{center}
\end{figure}

To explore the microwave properties of the BSCCO flake, we measured the temperature response of the resonance frequency on both the bare $\lambda/2$ resonator (R2) and the hybrid $\lambda/2$ + BSCCO resonator (R3), as shown in Figure \ref{fig3}.
For R2 (Figure \ref{fig3}a), the decrease in the resonance frequency with increased temperature is consistent with the thermal generation of equilibrium quasiparticles in a fully gapped superconductor $\left({\delta f}/{f_{0}}\right) \propto  \mathrm{e}^{-\Delta / T}$  \cite{mcdonald1987novel,prozorov2006magnetic} with a fitted superconducting gap of $\Delta_{Nb} = 1460\ \mathrm{\mu eV}$ that aligns with previous measurement of thin film niobium \cite{richards1960far,pronin1998direct}.
On the other hand, the $\lambda/2$ resonator with the BSCCO flake (R3) significantly deviates from this conventional behavior (see Figure \ref{fig3}b), showing instead a substantial upshift in the resonance frequency with increasing temperature. 
This behavior was consistently observed across several different BSCCO hybrid resonators (See Supporting Information Figure S6).
In superconducting resonators, a frequency upshift is often attributed to the interaction between the device and a two-level system (TLS) bath \cite{gao2008physics,mcrae2020materials,spiecker2023two}. 
The TLS bath is sometimes associated with charged imperfections in the dielectric material \cite{muller2019towards}. 
As the system's temperature increases the TLS bath saturates, leading to a suppression of the dispersive shift on the device and thus an upshift of the resonance frequency at higher temperature. 
The interaction with the TLS bath can be modeled by the following equation \cite{gao2008physics}:
\begin{align}
\begin{split}
\left(\frac{\delta f(T)}{f_{0}}\right)_{\mathrm{TLS}} =
\frac{1}{\pi Q_{\mathrm{TLS}}} \operatorname{Re}\biggl[\Psi\left(\frac{1}{2}+i \frac{\hbar \omega}{2 \pi k_{B} T}\right) 
-\ln \left(\frac{\hbar \omega}{2 \pi k_{B} T}\right)\biggr] 
\end{split}
\end{align}
where ${\Psi(x)}$ is the complex digamma function, and $Q_{\mathrm{TLS}}=1/F\tan(\delta)$ is the projected quality factor due to TLS loss which is proportional to the effective coupling $F$ and the material loss tangent $\tan(\delta)$.
We fit the resonance frequency shift of the hybrid resonator (R3) with a model that combines the TLS contribution and that of thermal quasiparticles:
\begin{align}
\frac{\delta f(T)}{f_{0}} &= \left(\frac{\delta f(T)}{f_{0}}\right)_{\mathrm{QP}} + \left(\frac{\delta f(T)}{f_{0}}\right)_{\mathrm{TLS}}
\end{align}
This model achieves good agreement with the data while introducing only a single additional fit parameter ($Q_{\mathrm{TLS}}$), and the fitted superconducting gap of the hybrid resonator R3 is consistent with the bare niobium resonators (R1, R2)
\begin{figure*}[!t]
\vspace{-0.5cm}
		\includegraphics[scale=0.3]{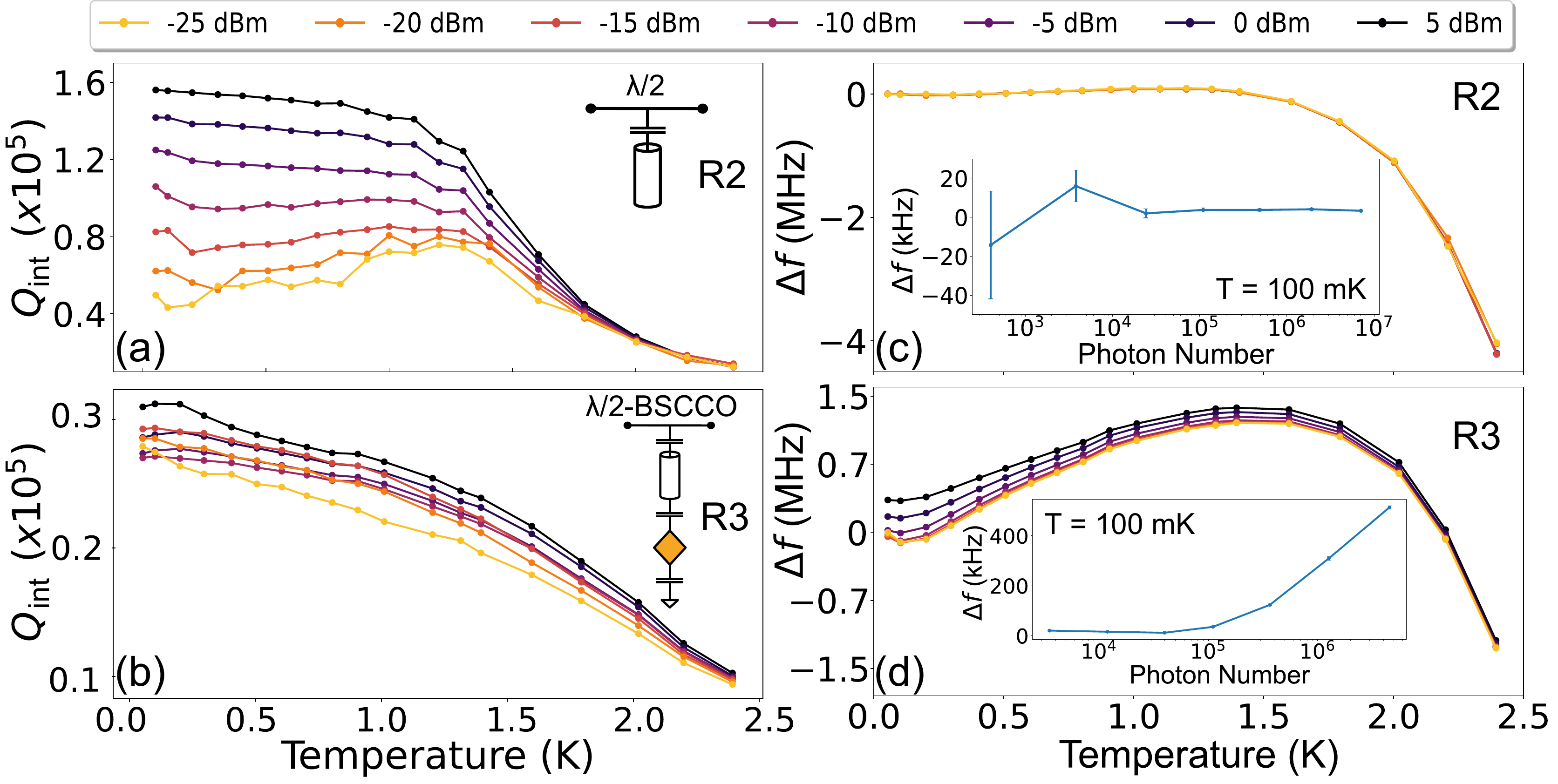}
        \caption{\textbf{Power dependence of the quality factor and resonance frequency}. (a-b) The internal quality factor $Q_{int}$ vs. temperature for different values of applied microwave power for resonator R2 (a) and resonator R3 (b). (c-d) Shift in the resonance frequency vs. temperature for different values of applied microwave power for resonator R2 (c) and resonator R3 (d). The inset shows frequency shift as a function of the applied microwave power at 100 mK, expressed in terms of the average intracavity photon number. }
		\label{fig4}
\end{figure*}
However, the quality factor we obtain from the fit model $Q_{\mathrm{TLS}} = 2300$  is one order of magnitude lower than the total internal quality factor of our device (R3).  
This discrepancy can be understood by considering a non-uniform spectral distribution of the TLS bath \cite{pappas2011two}. 
While TLS across a broad frequency range can contribute to a frequency shift, dissipation due to the TLS bath is dominated by resonant TLS with a frequency similar to the resonance frequency ($f_{\mathrm{TLS}} \approx f_{r}$). 
Thus, our data seems to suggest that the TLS bath in the BSCCO flake is mainly composed of off-resonant TLS. 

To further investigate the internal mechanisms within the BSCCO flake, we perform measurements on resonators R2 and R3 at varying drive powers. 
For the resonator R2 without the BSCCO flake, $Q_{int}$ improves by a factor of 4 with increasing power (see Figure \ref{fig4}a). 
At low temperatures, the quality factor shows weak temperature dependence, before decreasing rapidly due to the appearance of thermal quasiparticles.
This low-temperature power dependence is consistent with dielectric TLS loss in bare superconducting resonators \cite{altoe2022localization,Crowley2023disentangling,drimmer2024effect}, and matches the slight frequency upshift observed in Figure \ref{fig3}a.
For hybrid resonator R3, however, $Q_{int}$ appears power independent (see Figure \ref{fig4}b), indicating that it is limited by a different mechanism than the TLS in the bare niobium resonator. 
Additionally, $Q_{int}$ decreases with temperature with no visible plateau. While this can be caused by various effects, one potential explanation could be excess nodal quasiparticles excited even at low temperatures, due to the d-wave superconducting symmetry of BSCCO \cite{hosseini1999microwave,dahm2005nodal}.  

The resonance frequency power dependence also shows a distinct behavior for the hybrid circuit. 
While the bare resonator R2 is power independent within the measured power range ,shown in Figure \ref{fig4}c and its insert, the hybrid resonator shows a significant positive frequency shift (Figure \ref{fig4}d) and bifurcation consistent with Kerr nonlinearity. 
By estimating the number of photons in the resonator, we can obtain an effective Kerr nonlinearity of $K=0.1\ \mathrm{Hz/photon}$ at 100 mK (See Supporting Information). 
This value is comparable to those observed in high kinetic inductance resonators \cite{weissl2015kerr,maleeva2018circuit}, but notably, it is of the opposite sign. 
As the positive nonlinearity in Josephson devices requires precisely engineered multi-junction elements with external flux bias \cite{Frattini20173wave,zhang2017josephson,ranadive_kerr_2022}, the nonlinearity in the hybrid device is unlikely to be caused by Josephson or kinetic inductance effects.
Note that as the temperature increases, this nonlinearity diminishes, suggesting it is related to the saturation of the TLS bath.
However, a TLS bath typically has negligible effect on the power dependence of the resonator frequency \cite{gao2008physics,kumar2008temperature}. 
Recently, TLS-induced nonlinearity has been observed in several experiments in which an off-resonant pump tone was added to modify the spectral distribution of TLS defects \cite{kirsh2017revealing,capelle2020probing,andersson2021acoustic}.
Our observation of positive nonlinearity with a resonant frequency drive is thus additional evidence that the TLS bath of the hybrid device has an intrinsic non-uniform spectral density.
Another possibility is that the nonlinearity is due to nonequilibrium redistribution of nodal quasiparticles  \cite{fischer2023nonequilibrium}.

Our results show the unique behavior of saturable modes within the flake, but their microscopic origin is currently unknown. One contribution to the overall signal could come from the "pancake" layered vortex structure in BSCCO crystals.
Their short coherence length between the $\text{CuO}_{2}$ planes results in weak pinning \cite{blatter1994vortices}. 
Thus, at low temperatures it is possible that a single pancake pinning regime can be detected by the superconducting resonator \cite{niderost1996low}. 
However, given the finite size of the BSCCO flake in comparison to a BSCCO bulk crystal, it is also possible that effects from scale-free and intertwined lattice/charge/spin stripe inhomogeneities \cite{carlson2015decoding} could be detected. 
Another possible contribution could be due to interaction with mechanical modes in the flake \cite{will2017high,sahu2019nanoelectromechanical,sahu2022superconducting}, but it is unlikely as similar effects are observed for flakes of significantly different thickness (See Supporting Information Table S1).

In summary, we successfully integrated a BSCCO flake into a niobium co-planar resonator circuit by cryogenic transfer, demonstrating a significant advance in the coupling of high-temperature superconductors with superconducting circuits. 
The incorporation of BSCCO altered the resonant mode while maintaining a high-quality factor $(3\times 10^{4})$. 
Temperature-dependent measurements of the hybrid device revealed a significant upshift in the resonance frequency consistent with coupling to a TLS bath. However, the high quality factor of the resonator is inconsistent with strong coupling to such a system, indicating that the TLS bath is mainly off-resonant.
The hybrid device also exhibited significant positive nonlinearity, suggesting a new source of lossless nonlinearity unrelated to the Josephson effect. 
In addition to the novel observations of the BSCCO flake, this work shows a path towards high quality hybrid superconducting circuits with vdW materials and highlights their use in the exploration of unconventional superconductors and the development of new devices for quantum technology applications \cite{Brosco2024superconducting, Patel2024dmon, Coppo2024flux}. 

\section*{Acknowledgments}
The work is funded by the European Union (ERC-StG, cQEDscope, 101075962, ERC-CoG, 3DCuT, 101124606) and partially supported by the Deutsche Forschungsgemeinschaft (DFG 512734967, DFG 492704387, DFG 460444718, and DFG 452128813). The work at BNL was supported by the US Department of Energy, office of Basic Energy Sciences, contract no. DOE-sc0012704. BHG acknowledges additional support from Schmidt Science Fellows in partnership with Rhodes trust. The authors are  grateful to Ronny Engelhart, Heiko Reith, Christiane Kranz, and Martin Bauer for technical support, and to Valentina Brosco, Bernd B{\"u}chner, Gianluigi Catelani, Luca Chiroli, Mathieu Fechant, Ioan Pop, and Martin Spiecker for fruitful discussions.

\section*{references}

\begin{thebibliography}{}

\bibitem{du2008approaching}Du, X., Skachko, I., Barker, A. \& Andrei, E. Approaching ballistic transport in suspended graphene. {\em Nature Nanotechnology}. \textbf{3}, 491-495 (2008)
\bibitem{mayorov2011micrometer}Mayorov, A., Gorbachev, R., Morozov, S., Britnell, L., Jalil, R., Ponomarenko, L., Blake, P., Novoselov, K., Watanabe, K., Taniguchi, T. \& Others Micrometer-scale ballistic transport in encapsulated graphene at room temperature. {\em Nano Letters}. \textbf{11}, 2396-2399 (2011)
\bibitem{gong2017discovery}Gong, C., Li, L., Li, Z., Ji, H., Stern, A., Xia, Y., Cao, T., Bao, W., Wang, C., Wang, Y. \& Others Discovery of intrinsic ferromagnetism in two-dimensional van der Waals crystals. {\em Nature}. \textbf{546}, 265-269 (2017)
\bibitem{huang2018electrical}Huang, B., Clark, G., Klein, D., MacNeill, D., Navarro-Moratalla, E., Seyler, K., Wilson, N., McGuire, M., Cobden, D., Xiao, D. \& Others Electrical control of 2D magnetism in bilayer Cr$\text I_{3}$. {\em Nature Nanotechnology}. \textbf{13}, 544-548 (2018)
\bibitem{wang2022magnetic}Wang, Q., Bedoya-Pinto, A., Blei, M., Dismukes, A., Hamo, A., Jenkins, S., Koperski, M., Liu, Y., Sun, Q., Telford, E. \& Others The magnetic genome of two-dimensional van der Waals materials. {\em ACS Nano}. \textbf{16}, 6960-7079 (2022)
\bibitem{deng2020quantum}Deng, Y., Yu, Y., Shi, M., Guo, Z., Xu, Z., Wang, J., Chen, X. \& Zhang, Y. Quantum anomalous Hall effect in intrinsic magnetic topological insulator Mn$\text{Bi}_{2}\text{Te}_{4}$. {\em Science}. \textbf{367}, 895-900 (2020)
\bibitem{wu2018observation}Wu, S., Fatemi, V., Gibson, Q., Watanabe, K., Taniguchi, T., Cava, R. \& Jarillo-Herrero, P. Observation of the quantum spin Hall effect up to 100 kelvin in a monolayer crystal. {\em Science}. \textbf{359}, 76-79 (2018)
\bibitem{fatemi2018electrically}Fatemi, V., Wu, S., Cao, Y., Bretheau, L., Gibson, Q., Watanabe, K., Taniguchi, T., Cava, R. \& Jarillo-Herrero, P. Electrically tunable low-density superconductivity in a monolayer topological insulator. {\em Science}. \textbf{362}, 926-929 (2018)
\bibitem{xi2016ising}Xi, X., Wang, Z., Zhao, W., Park, J., Law, K., Berger, H., Forró, L., Shan, J. \& Mak, K. Ising pairing in superconducting Nb$\text Se_2$ atomic layers. {\em Nature Physics}. \textbf{12}, 139-143 (2016) 
\bibitem{zhao2019sign}Zhao, S., Poccia, N., Panetta, M., Yu, C., Johnson, J., Yoo, H., Zhong, R., Gu, G., Watanabe, K., Taniguchi, T. \& Others Sign-reversing Hall effect in atomically thin high-temperature $\text {Bi}_{2.1}\text{Sr}_{1.9}\text{CaCu}_{2.0}\text{O}_{8+\delta}$ superconductors. {\em Physical Review Letters}. \textbf{122}, 247001 (2019)
\bibitem{yu2019high}Yu, Y., Ma, L., Cai, P., Zhong, R., Ye, C., Shen, J., Gu, G., Chen, X. \& Zhang, Y. High-temperature superconductivity in monolayer $\text{Bi}_{2}\text{Sr}_{2}\text{CaCu}_{2}\text{O}_{8+\delta}$. {\em Nature}. \textbf{575}, 156-163 (2019)
\bibitem{meng2024layer}Meng, K., Zhang, X., Song, B., Li, B., Kong, X., Huang, S., Yang, X., Jin, X., Wu, Y., Nie, J. \& Others Layer-Dependent Superconductivity in Iron-Based Superconductors $\text{Ca}_2\text{Fe}_4\text{As}_4\text{F}_2$ and CaK$\text{Fe}_4\text{As}_4$. {\em Nano Letters}.  \textbf{24}, 6821–6827 (2024)
\bibitem{cao2018unconventional}Cao, Y., Fatemi, V., Fang, S., Watanabe, K., Taniguchi, T., Kaxiras, E. \& Jarillo-Herrero, P. Unconventional superconductivity in magic-angle graphene superlattices. {\em Nature}. \textbf{556}, 43-50 (2018)
\bibitem{kang2024evidence}Kang, K., Shen, B., Qiu, Y., Zeng, Y., Xia, Z., Watanabe, K., Taniguchi, T., Shan, J. \& Mak, K. Evidence of the fractional quantum spin Hall effect in moiré Mo$\text{Te}_2$. {\em Nature}. \textbf{628}, 522-526 (2024)
\bibitem{hosseini1999microwave}Hosseini, A., Harris, R., Kamal, S., Dosanjh, P., Preston, J., Liang, R., Hardy, W. \& Bonn, D. Microwave spectroscopy of thermally excited quasiparticles in Y$\text{Ba}_2\text{Cu}_3\text{O}_{6.99}$. {\em Physical Review B}. \textbf{60}, 1349 (1999)
\bibitem{prozorov2006magnetic}Prozorov, R. \& Giannetta, R. Magnetic penetration depth in unconventional superconductors. {\em Superconductor Science And Technology}. \textbf{19}, R41 (2006)
\bibitem{thiemann2018single}Thiemann, M., Beutel, M., Dressel, M., Lee-Hone, N., Broun, D., Fillis-Tsirakis, E., Boschker, H., Mannhart, J. \& Scheffler, M. Single-Gap Superconductivity and Dome of Superfluid Density in Nb-Doped SrTi$\text{O}_3$. {\em Physical Review Letters}. \textbf{120}, 237002 (2018)
\bibitem{Phan2022detecting}Phan, D., Senior, J., Ghazaryan, A., Hatefipour, M., Strickland, W., Shabani, J., Serbyn, M. \& Higginbotham, A. Detecting Induced p+ip Pairing at the Al-InAs Interface with a Quantum Microwave Circuit. {\em Phys. Rev. Lett.}. \textbf{128}, 107701 (2022)
\bibitem{bottcher2023circuit}Bøttcher, C., Poniatowski, N., Grankin, A., Wesson, M., Yan, Z., Vool, U., Galitski, V. \& Yacoby, A. Circuit quantum electrodynamics detection of induced two-fold anisotropic pairing in a hybrid superconductor–ferromagnet bilayer. {\em Nature Physics}. \textbf{20}, 1609–1615 (2024)
\bibitem{hammer2008ultra}Hammer, G., Wuensch, S., Ilin, K. \& Siegel, M. Ultra high quality factor resonators for kinetic inductance detectors. {\em Journal Of Physics: Conference Series}. \textbf{97}, 012044 (2008)
\bibitem{gao2008physics}Gao, J. The physics of superconducting microwave resonators. California Institute of Technology, 192 (2008)
\bibitem{megrant2012planar}Megrant, A., Neill, C., Barends, R., Chiaro, B., Chen, Y., Feigl, L., Kelly, J., Lucero, E., Mariantoni, M., O’Malley, P. \& Others Planar superconducting resonators with internal quality factors above one million. {\em Applied Physics Letters}. \textbf{100}, 113510 (2012)
\bibitem{mahashabde2020fast}Mahashabde, S., Otto, E., Montemurro, D., Graaf, S., Kubatkin, S. \& Danilov, A. Fast tunable high-q-factor superconducting microwave resonators. {\em Physical Review Applied}. \textbf{14}, 044040 (2020)
\bibitem{krantz2019quantum}Krantz, P., Kjaergaard, M., Yan, F., Orlando, T., Gustavsson, S. \& Oliver, W. A quantum engineer's guide to superconducting qubits. {\em Applied Physics Reviews}. \textbf{6}, 021318 (2019)
\bibitem{blais2021circuit}Blais, A., Grimsmo, A., Girvin, S. \& Wallraff, A. Circuit quantum electrodynamics. {\em Reviews Of Modern Physics}. \textbf{93}, 025005 (2021)
\bibitem{antony2021making}Antony, A., Gustafsson, M., Rajendran, A., Benyamini, A., Ribeill, G., Ohki, T., Hone, J. \& Fong, K. Making high-quality quantum microwave devices with van der Waals superconductors. {\em Journal Of Physics: Condensed Matter}. \textbf{34}, 103001 (2021)
\bibitem{wang2022hexagonal}Wang, J., Yamoah, M., Li, Q., Karamlou, A., Dinh, T., Kannan, B., Braumüller, J., Kim, D., Melville, A., Muschinske, S. \& Others Hexagonal boron nitride as a low-loss dielectric for superconducting quantum circuits and qubits. {\em Nature Materials}. \textbf{21}, 398-403 (2022)
\bibitem{maji2024superconducting}Maji, K., Sarkar, J., Mandal, S., Hingankar, M., Mukherjee, A., Samal, S., Bhattacharjee, A., Patankar, M., Watanabe, K., Taniguchi, T. \& Others Superconducting Cavity-Based Sensing of Band Gaps in 2D Materials. {\em Nano Letters}. \textbf{24}, 4369-4375 (2024)
\bibitem{kreidel2024measuring}Kreidel, M., Chu, X., Balgley, J., Antony, A., Verma, N., Ingham, J., Ranzani, L., Queiroz, R., Westervelt, R., Hone, J. \& Others Measuring kinetic inductance and superfluid stiffness of two-dimensional superconductors using high-quality transmission-line resonators. {\em Physical Review Research}. \textbf{6}, 043245 (2024)
\bibitem{tanaka2024superfluid}Tanaka, M., Î-Wang, J., Dinh, T., Rodan-Legrain, D., Zaman, S., Hays, M., Kannan, B., Almanakly, A., Kim, D., Niedzielski, B., Serniak, K., Schwartz, M., Watanabe, K., Taniguchi, T., Grover, J., Orlando, T., Gustavsson, S., Jarillo-Herrero, P. \& Oliver, W. Superfluid Stiffness and Flat-Band Superconductivity in Magic-Angle Graphene Probed by cQED. arXiv:2406.13740 2024-10-31 https://arxiv.org/abs/2406.13740

\bibitem{schmidt2018ballistic}Schmidt, F., Jenkins, M., Watanabe, K., Taniguchi, T. \& Steele, G. A ballistic graphene superconducting microwave circuit. {\em Nature Communications}. \textbf{9}, 4069 (2018)
\bibitem{wang2019coherent}Wang, J., Rodan-Legrain, D., Bretheau, L., Campbell, D., Kannan, B., Kim, D., Kjaergaard, M., Krantz, P., Samach, G., Yan, F. \& Others Coherent control of a hybrid superconducting circuit made with graphene-based van der Waals heterostructures. {\em Nature Nanotechnology}. \textbf{14}, 120-125 (2019)
\bibitem{haller2022phase}Haller, R., Fülöp, G., Indolese, D., Ridderbos, J., Kraft, R., Cheung, L., Ungerer, J., Watanabe, K., Taniguchi, T., Beckmann, D. \& Others Phase-dependent microwave response of a graphene Josephson junction. {\em Physical Review Research}. \textbf{4}, 013198 (2022)
\bibitem{butseraen2022gate}Butseraen, G., Ranadive, A., Aparicio, N., Rafsanjani Amin, K., Juyal, A., Esposito, M., Watanabe, K., Taniguchi, T., Roch, N., Lefloch, F. \& Others A gate-tunable graphene Josephson parametric amplifier. {\em Nature Nanotechnology}. \textbf{17}, 1153-1158 (2022)
\bibitem{zhao2023time}Zhao, S., Cui, X., Volkov, P., Yoo, H., Lee, S., Gardener, J., Akey, A., Engelke, R., Ronen, Y., Zhong, R. \& Others Time-reversal symmetry breaking superconductivity between twisted cuprate superconductors. {\em Science}. \textbf{382}, 1422-1427 (2023)
\bibitem{lee2023encapsulating}Lee, Y., Martini, M., Confalone, T., Shokri, S., Saggau, C., Wolf, D., Gu, G., Watanabe, K., Taniguchi, T., Montemurro, D. \& Others Encapsulating High-Temperature Superconducting Twisted van der Waals Heterostructures Blocks Detrimental Effects of Disorder. {\em Advanced Materials}. \textbf{35}, 2209135 (2023)
\bibitem{martini2023twisted}Martini, M., Lee, Y., Confalone, T., Shokri, S., Saggau, C., Wolf, D., Gu, G., Watanabe, K., Taniguchi, T., Montemurro, D. \& Others Twisted cuprate van der Waals heterostructures with controlled Josephson coupling. {\em Materials Today}. \textbf{67}, 106-112 (2023)
\bibitem{can2021high}Can, O., Tummuru, T., Day, R., Elfimov, I., Damascelli, A. \& Franz, M. High-temperature topological superconductivity in twisted double-layer copper oxides. {\em Nature Physics}. \textbf{17}, 519-524 (2021)
\bibitem{song2022doping}Song, X., Zhang, Y. \& Vishwanath, A. Doping a moiré Mott insulator: A t\ensuremath-J model study of twisted cuprates. {\em Phys. Rev. B}. \textbf{105}, L201102 (2022)
\bibitem{liu2023charge}Liu, Y., Zhou, J., Wu, C. \& Yang, F. Charge-4e superconductivity and chiral metal in 45°-twisted bilayer cuprates and related bilayers. {\em Nature Communications}. \textbf{14}, 7926 (2023)
\bibitem{yuan2023inhomogeneity}Yuan, A., Vituri, Y., Berg, E., Spivak, B. \& Kivelson, S. Inhomogeneity-induced time-reversal symmetry breaking in cuprate twist junctions. {\em Physical Review B}. \textbf{108}, L100505 (2023)
\bibitem{talantsev2024evidences}Talantsev, E. Evidences for d-wave symmetry of c-axis superconducting gap in atomically thin twisted flakes of bismuth-based HTS cuprates. {\em Physica C: Superconductivity And Its Applications}. \textbf{635}, 1354549 (2024)
\bibitem{zeljkovic2012imaging}Zeljkovic, I., Xu, Z., Wen, J., Gu, G., Markiewicz, R. \& Hoffman, J. Imaging the impact of single oxygen atoms on superconducting $\text {Bi}_{2+y}\text{Sr}_{2-y}\text{CaCu}_{2}\text{O}_{8+x}$. {\em Science}. \textbf{337}, 320-323 (2012)
\bibitem{poccia2020spatially}Poccia, N., Zhao, S., Yoo, H., Huang, X., Yan, H., Chu, Y., Zhong, R., Gu, G., Mazzoli, C., Watanabe, K. \& Others Spatially correlated incommensurate lattice modulations in an atomically thin high-temperature $\text {Bi}_{2.1}\text{Sr}_{1.9}\text{CaCu}_{2.0}\text{O}_{8+y}$ superconductor. {\em Physical Review Materials}. \textbf{4}, 114007 (2020)
\bibitem{figueruelo2024apparent}Figueruelo-Campanero, I., Campo, A., Nieva, G., González, E., Serrano, A. \& Menghini, M. Apparent color and Raman vibrational modes of the high-temperature superconductor $\text {Bi}_{2}\text{Sr}_{2}\text{CaCu}_{2}\text{O}_{8+\delta}$  exfoliated flakes. {\em 2D Materials}. \textbf{11}, 025032 (2024)
\bibitem{sahu2019nanoelectromechanical}Sahu, S., Vaidya, J., Schmidt, F., Jangade, D., Thamizhavel, A., Steele, G., Deshmukh, M. \& Singh, V. Nanoelectromechanical resonators from high-Tc superconducting crystals of $\text {Bi}_{2}\text{Sr}_{2}\text{Ca}_1\text{Cu}_{2}\text{O}_{8+\delta}$. {\em 2D Materials}. \textbf{6}, 025027 (2019)
\bibitem{ghosh2020demand}Ghosh, S., Vaidya, J., Datta, S., Pandeya, R., Jangade, D., Kulkarni, R., Maiti, K., Thamizhavel, A. \& Deshmukh, M. On-Demand Local Modification of High-Tc Superconductivity in Few Unit-Cell Thick $\text {Bi}_{2}\text{Sr}_{2}\text{CaCu}_{2}\text{O}_{8+\delta}$. {\em Advanced Materials}. \textbf{32}, 2002220 (2020)
\bibitem{mcdonald1987novel}McDonald, D. Novel superconducting thermometer for bolometric applications. {\em Applied Physics Letters}. \textbf{50}, 775-777 (1987)
\bibitem{richards1960far}Richards, P. \& Tinkham, M. Far-infrared energy gap measurements in bulk superconducting In, Sn, Hg, Ta, V, Pb and Nb. {\em Physical Review}. \textbf{119}, 575 (1960)
\bibitem{pronin1998direct}Pronin, A., Dressel, M., Pimenov, A., Loidl, A., Roshchin, I. \& Greene, L. Direct observation of the superconducting energy gap developing in the conductivity spectra of niobium. {\em Physical Review B}. \textbf{57}, 14416 (1998)
\bibitem{mcrae2020materials}McRae, C., Wang, H., Gao, J., Vissers, M., Brecht, T., Dunsworth, A., Pappas, D. \& Mutus, J. Materials loss measurements using superconducting microwave resonators. {\em Review Of Scientific Instruments}. \textbf{91} (2020)
\bibitem{spiecker2023two}Spiecker, M., Paluch, P., Gosling, N., Drucker, N., Matityahu, S., Gusenkova, D., Günzler, S., Rieger, D., Takmakov, I., Valenti, F. \& Others Two-level system hyperpolarization using a quantum Szilard engine. {\em Nature Physics}. \textbf{19}, 1320-1325 (2023)
\bibitem{muller2019towards}Müller, C., Cole, J. \& Lisenfeld, J. Towards understanding two-level-systems in amorphous solids: insights from quantum circuits. {\em Reports On Progress In Physics}. \textbf{82}, 124501 (2019)
\bibitem{pappas2011two}Pappas, D., Vissers, M., Wisbey, D., Kline, J. \& Gao, J. Two level system loss in superconducting microwave resonators. {\em IEEE Transactions On Applied Superconductivity}. \textbf{21}, 871-874 (2011)
\bibitem{altoe2022localization}Altoé, M., Banerjee, A., Berk, C., Hajr, A., Schwartzberg, A., Song, C., Alghadeer, M., Aloni, S., Elowson, M., Kreikebaum, J. \& Others Localization and mitigation of loss in niobium superconducting circuits. {\em PRX Quantum}. \textbf{3}, 020312 (2022)
\bibitem{Crowley2023disentangling}Crowley, K., McLellan, R., Dutta, A., Shumiya, N., Place, A., Le, X., Gang, Y., Madhavan, T., Bland, M., Chang, R., Khedkar, N., Feng, Y., Umbarkar, E., Gui, X., Rodgers, L., Jia, Y., Feldman, M., Lyon, S., Liu, M., Cava, R., Houck, A. \& Leon, N. Disentangling Losses in Tantalum Superconducting Circuits. {\em Phys. Rev. X}. \textbf{13}, 041005 (2023)
\bibitem{drimmer2024effect}Drimmer, M., Telkamp, S., Fischer, F., Rodrigues, I., Todt, C., Krizek, F., Kriegner, D., Müller, C., Wegscheider, W. \& Chu, Y. The effect of niobium thin film structure on losses in superconducting circuits. arXiv:2403.12164 2024-03-18 https://arxiv.org/abs/2403.12164
\bibitem{dahm2005nodal}Dahm, T., Hirschfeld, P., Scalapino, D. \& Zhu, L. Nodal quasiparticle lifetimes in cuprate superconductors. {\em Physical Review B}. \textbf{72}, 214512 (2005)
\bibitem{weissl2015kerr}Weißl, T., Küng, B., Dumur, É., Feofanov, A., Matei, I., Naud, C., Buisson, O., Hekking, F. \& Guichard, W. Kerr coefficients of plasma resonances in Josephson junction chains. {\em Physical Review B}. \textbf{92}, 104508 (2015)
\bibitem{maleeva2018circuit}Maleeva, N., Grünhaupt, L., Klein, T., Levy-Bertrand, F., Dupre, O., Calvo, M., Valenti, F., Winkel, P., Friedrich, F., Wernsdorfer, W. \& Others Circuit quantum electrodynamics of granular aluminum resonators. {\em Nature Communications}. \textbf{9}, 3889 (2018)
\bibitem{Frattini20173wave}Frattini, N., Vool, U., Shankar, S., Narla, A., Sliwa, K. \& Devoret, M. 3-wave mixing Josephson dipole element. {\em Applied Physics Letters}. \textbf{110}, 222603 (2017)
\bibitem{zhang2017josephson}Zhang, W., Huang, W., Gershenson, M. \& Bell, M. Josephson Metamaterial with a Widely Tunable Positive or Negative Kerr Constant. {\em Phys. Rev. Appl.}. \textbf{8}, 051001 (2017,11)
\bibitem{ranadive_kerr_2022}Ranadive, A., Esposito, M., Planat, L., Bonet, E., Naud, C., Buisson, O., Guichard, W. \& Roch, N. Kerr reversal in Josephson meta-material and traveling wave parametric amplification. {\em Nature Communications}. \textbf{13}, 1737 (2022)
\bibitem{kumar2008temperature}Kumar, S., Gao, J., Zmuidzinas, J., Mazin, B., LeDuc, H. \& Day, P. Temperature dependence of the frequency and noise of superconducting coplanar waveguide resonators. {\em Applied Physics Letters}. \textbf{92}, 123503 (2008)
\bibitem{kirsh2017revealing}Kirsh, N., Svetitsky, E., Burin, A., Schechter, M. \& Katz, N. Revealing the nonlinear response of a tunneling two-level system ensemble using coupled modes. {\em Physical Review Materials}. \textbf{1}, 012601 (2017)
\bibitem{capelle2020probing}Capelle, T., Flurin, E., Ivanov, E., Palomo, J., Rosticher, M., Chua, S., Briant, T., Cohadon, P., Heidmann, A., Jacqmin, T. \& Others Probing a two-level system bath via the frequency shift of an off-resonantly driven cavity. {\em Physical Review Applied}. \textbf{13}, 034022 (2020)
\bibitem{andersson2021acoustic}Andersson, G., Bilobran, A., Scigliuzzo, M., Lima, M., Cole, J. \& Delsing, P. Acoustic spectral hole-burning in a two-level system ensemble. {\em Npj Quantum Information}. \textbf{7}, 15 (2021)
\bibitem{fischer2023nonequilibrium}Fischer, P. \& Catelani, G. Nonequilibrium quasiparticle distribution in superconducting resonators: An analytical approach. {\em Physical Review Applied}. \textbf{19}, 054087 (2023)
\bibitem{blatter1994vortices}Blatter, G., Feigel'man, M., Geshkenbein, V., Larkin, A. \& Vinokur, V. Vortices in high-temperature superconductors. {\em Reviews Of Modern Physics}. \textbf{66}, 1125 (1994)
\bibitem{niderost1996low}Nideröst, M., Suter, A., Visani, P., Mota, A. \& Blatter, G. Low-field vortex dynamics over seven time decades in a $\text {Bi}_{2}\text{Sr}_{2}\text{CaCu}_{2}\text{O}_{8+\delta}$ single crystal for temperatures 13 $< \sim T < \sim 83$ K. {\em Physical Review B}. \textbf{53}, 9286 (1996)
\bibitem{carlson2015decoding}Carlson, E., Liu, S., Phillabaum, B. \& Dahmen, K. Decoding spatial complexity in strongly correlated electronic systems. {\em Journal Of Superconductivity And Novel Magnetism}. \textbf{28}, 1237-1243 (2015)
\bibitem{will2017high}Will, M., Hamer, M., Muller, M., Noury, A., Weber, P., Bachtold, A., Gorbachev, R., Stampfer, C. \& Guttinger, J. High quality factor graphene-based two-dimensional heterostructure mechanical resonator. {\em Nano Letters}. \textbf{17}, 5950-5955 (2017)
\bibitem{sahu2022superconducting}Sahu, S., Mandal, S., Ghosh, S., Deshmukh, M. \& Singh, V. Superconducting vortex-charge measurement using cavity electromechanics. {\em Nano Letters}. \textbf{22}, 1665-1671 (2022)
\bibitem{Brosco2024superconducting}Brosco, V., Serpico, G., Vinokur, V., Poccia, N. \& Vool, U. Superconducting Qubit Based on Twisted Cuprate Van der Waals Heterostructures. {\em Phys. Rev. Lett.}. \textbf{132}, 017003 (2024), 
\bibitem{Patel2024dmon}Patel, H., Pathak, V., Can, O., Potter, A. \& Franz, M. d-Mon: A Transmon with Strong Anharmonicity Based on Planar c-Axis Tunneling Junction between d-Wave and s-Wave Superconductors. {\em Phys. Rev. Lett.}. \textbf{132}, 017002 (2024)
\bibitem{Coppo2024flux}Coppo, A., Chirolli, L., Poccia, N., Vool, U. \& Brosco, V. Flux-Tunable Regimes and Supersymmetry in Twisted Cuprate Heterostructures. {\em Applied Physics Letters}. \textbf{125}, 054001 (2024)







\end{thebibliography}
{}
\end{document}


\newpage
\setcounter{secnumdepth}{2}

\onecolumngrid
\appendix
\begin{center}
\large{\textbf{Supporting Information for \\ ``Exploring van der Waals cuprate superconductors using a hybrid microwave circuit''}}
\end{center}

\author{Haolin Jin}
\thanks{These authors contributed equally to this work.}
\affiliation{Max Planck Institute for Chemical Physics of Solids, 01187 Dresden, Germany}
\affiliation{Institute of Solid State and Material Physics, Technische Universit{\"a}t Dresden, 01062 Dresden, Germany}

\noindent

\author{Giuseppe Serpico}
\thanks{These authors contributed equally to this work.}
\affiliation{Max Planck Institute for Chemical Physics of Solids, 01187 Dresden, Germany}
\affiliation{Department of Physics, University of Naples Federico II, Via Cintia, 80126 Naples, Italy}

\author{Yejin Lee}
\affiliation{Max Planck Institute for Chemical Physics of Solids, 01187 Dresden, Germany}

\author{Tommaso Confalone}
\affiliation{Leibniz Institute for Solid State and Materials Science Dresden (IFW Dresden), 01069 Dresden, Germany}
\affiliation{Institute of Applied Physics, Technische Universit{\"a}t Dresden, 01062 Dresden, Germany}

\author{Christian N. Saggau}
\affiliation{Leibniz Institute for Solid State and Materials Science Dresden (IFW Dresden), 01069 Dresden, Germany}
 \affiliation{DTU Electro, Department of Electrical and Photonics Engineering, Technical University of Denmark, 2800 Kongens Lyngby, Denmark}
\affiliation{Center for Silicon Photonics for Optical Communications (SPOC), Technical University of Denmark, 2800 Kongens Lyngby, Denmark}

\author{Flavia Lo Sardo}
\affiliation{Leibniz Institute for Solid State and Materials Science Dresden (IFW Dresden), 01069 Dresden, Germany}
\affiliation{Institute of Materials Science, Technische Universit{\"a}t Dresden, 01062 Dresden, Germany}

\author{Genda Gu}
\affiliation{Condensed Matter Physics and Materials Science Department, Brookhaven National Laboratory, Upton, NY 11973, USA}

\author{Berit H. Goodge}
\affiliation{Max Planck Institute for Chemical Physics of Solids, 01187 Dresden, Germany}

\author{Edouard Lesne}
\affiliation{Max Planck Institute for Chemical Physics of Solids, 01187 Dresden, Germany}

\author{Domenico Montemurro}
\affiliation{Department of Physics, University of Naples Federico II, Via Cintia, 80126 Naples, Italy}

\author{Kornelius Nielsch}
\affiliation{Leibniz Institute for Solid State and Materials Science Dresden (IFW Dresden), 01069 Dresden, Germany}
\affiliation{Institute of Applied Physics, Technische Universit{\"a}t Dresden, 01062 Dresden, Germany}
\affiliation{Institute of Materials Science, Technische Universit{\"a}t Dresden, 01062 Dresden, Germany}

\author{Nicola Poccia}
\affiliation{Department of Physics, University of Naples Federico II, Via Cintia, 80126 Naples, Italy}
\affiliation{Leibniz Institute for Solid State and Materials Science Dresden (IFW Dresden), 01069 Dresden, Germany}

\author{Uri Vool}
\thanks{uri.vool@cpfs.mpg.de}
\affiliation{Max Planck Institute for Chemical Physics of Solids, 01187 Dresden, Germany}
\affiliation{Leibniz Institute for Solid State and Materials Science Dresden (IFW Dresden), 01069 Dresden, Germany}

\maketitle
\onecolumngrid
\makeatletter
\renewcommand{\fnum@figure}{\textbf{Figure~\thefigure}}
\makeatother
\renewcommand{\thefigure}{S\arabic{figure}}
\newpage
\renewcommand{\thetable}{S\arabic{table}}\makeatletter
\renewcommand{\fnum@table}{\textbf{Table~\thetable}}
\makeatother

\section{Coupling capacitance simulation}
\begin{figure*}[!h]
	\begin{center}
		\includegraphics[scale=0.45]{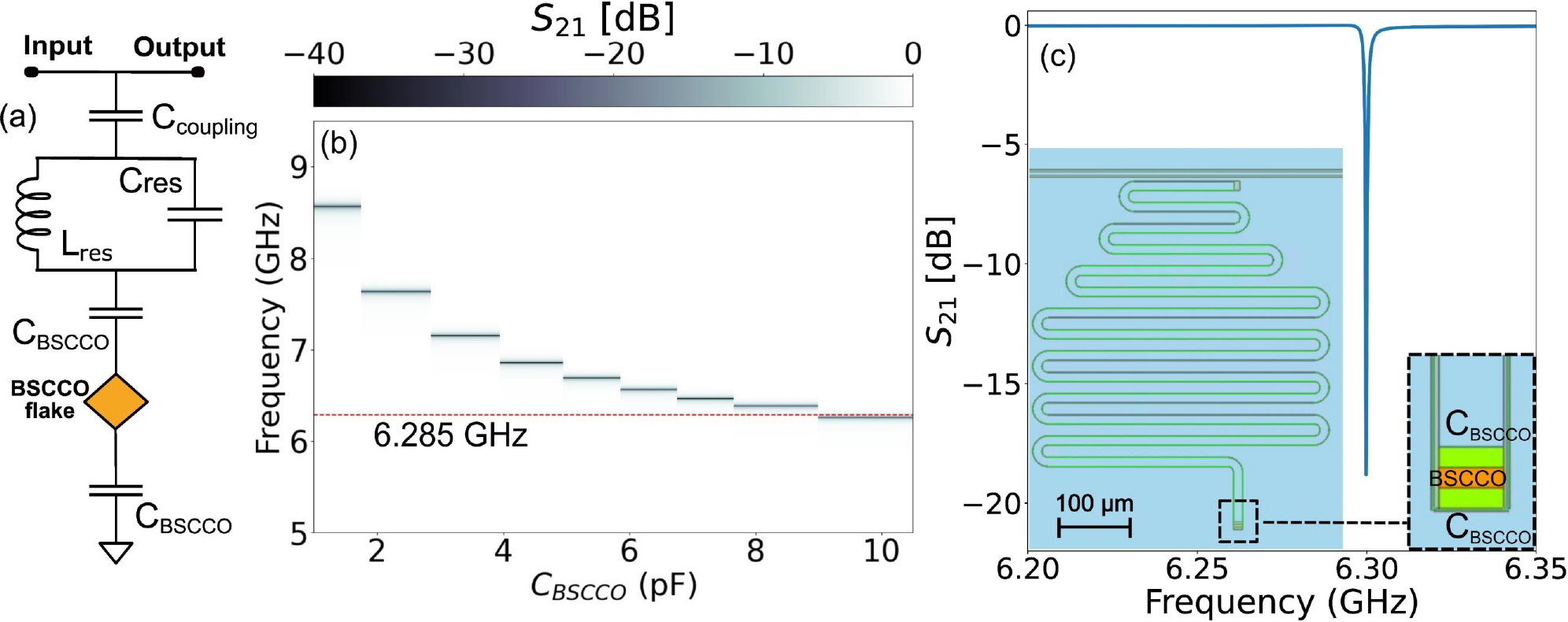}
        \caption{\textbf{BSCCO capacitance simulation}.  (a) Schematic of the circuit representing a superconducting resonator coupled to a BSCCO flake through $C_{\text{BSCCO}}$. (b) Simulation results of the $S_{21}$ parameters using QUCS for different values of $C_{\text{BSCCO}}$. The red line indicates the experimentally measured value of the resonance frequency. (c) HFSS-Ansys simulation of the circuit, providing a more accurate model of the system. The simulation results suggest a coupling capacitance of approximately 5 pF. The inset shows a view of the coupling region between the resonator and the BSCCO flake.}
		\label{fig5}
	\end{center}
\end{figure*}

When the BSCCO is coupled to the resonator, it is expected to effectively short the circuit, transitioning its boundary conditions from half wavelength to quarter wavelength.
However, experimental measurements reveal a resonance frequency of 6.285 GHz, which deviates from the anticipated value of 5.264 GHz. 
This discrepancy suggests that the BSCCO flake should not be treated as galvanically coupled to the resonator, and its coupling can be modelled by having a 
 finite capacitance between the resonator and the flake. 
To investigate this quantitatively, we adopted a circuit model, as illustrated in Figure \ref{fig5}(a), where the BSCCO is treated as an inductance and the coupling to the resonator and ground plane are treated as two equal capacitors $C_{\text{BSCCO}}$.
Initially, to estimate the capacitance, we conducted simulations using a circuit simulator (Qucsstudio), performing S-parameter simulations for capacitance values ranging from 100 fF to 10 pF. 
The results depicted in Figure \ref{fig5}(b) illustrate the variation in resonance frequency with changes in the capacitive coupling of the BSCCO.
Around 9 pF, the simulation yields a resonance frequency value consistent with experimental data.
Subsequently, a more detailed 3-dimensional S-parameter simulation was performed using HFSS-Ansys (Figure \ref{fig5}(c)). This simulation revealed that the coupling capacitance is approximately 5 pF, providing a more accurate estimation of the BSCCO resonator coupling characteristics.

\section{Fabrication of niobium resonators}
The niobium thin films were sputtered on pure intrinsic Si <100> single-crystal substrates using a BESTEC ultra high vacuum magnetron sputtering system. 
Before deposition, the chamber was evacuated to a base pressure of less than $8 \times 10^{-9}$ mbar, while the process gas (Ar 5 N) pressure was set to  $ 3 \times 10^{-3}$ mbar. 
The target-to-substrate distance was fixed at 20 cm, and the substrate was rotated during the deposition to ensure a homogeneous growth. 
The sputtering process took 40 mins for a 60 nm thick film.
To avoid oxidation of the films, an amorphous Si layer of 6 nm was deposited in situ. The entire process was done at room temperature. 

To pattern the resonator, SML300 EBL resist was spun on the film at 2000 rpm for 40 secs with a ram rate of 1000 rpm/sec for an approximate resist thickness of 500 nm and soft baked at $\SI{180}\celsius$ for 5 minute. 
The EBL resist was patterned using a VOYAGER (Raith nanofabrication, Germany).
The resist was developed in de-ionized water/isopropanol (3/7) solution for 60 s and rinsed in isopropanol for 10 s. 
After development, the mask was hard baked at $\SI{80}\celsius$ for 30 minutes. 
Using the patterned resist as a mask, we etched with the niobium using a dry etch technique. 
The etching parameters we used were: ambient pressure 0.015 mbar, $\text{CHF}_{3}$ flow rate 20 sccm, $\text{CF}_{4}$ flow rate 10 sccm, $\text{O}_{2}$ flow rate 3 sccm, He flow rate 7 sccm, ICP power of 250 W, and RF bias power of 15 W at a temperature of $\SI{30}\celsius$.
After etching, the resist mask was removed in Acetone with 10 min sonication followed by rinsing in isopropanol. 
Finally, $\text{O}_{2}$ plasma was applied for 10 min to remove the residual polymers.

\section{Preparation of the hybrid device}

\begin{figure*}[!h]
    \includegraphics[width=2.8in]{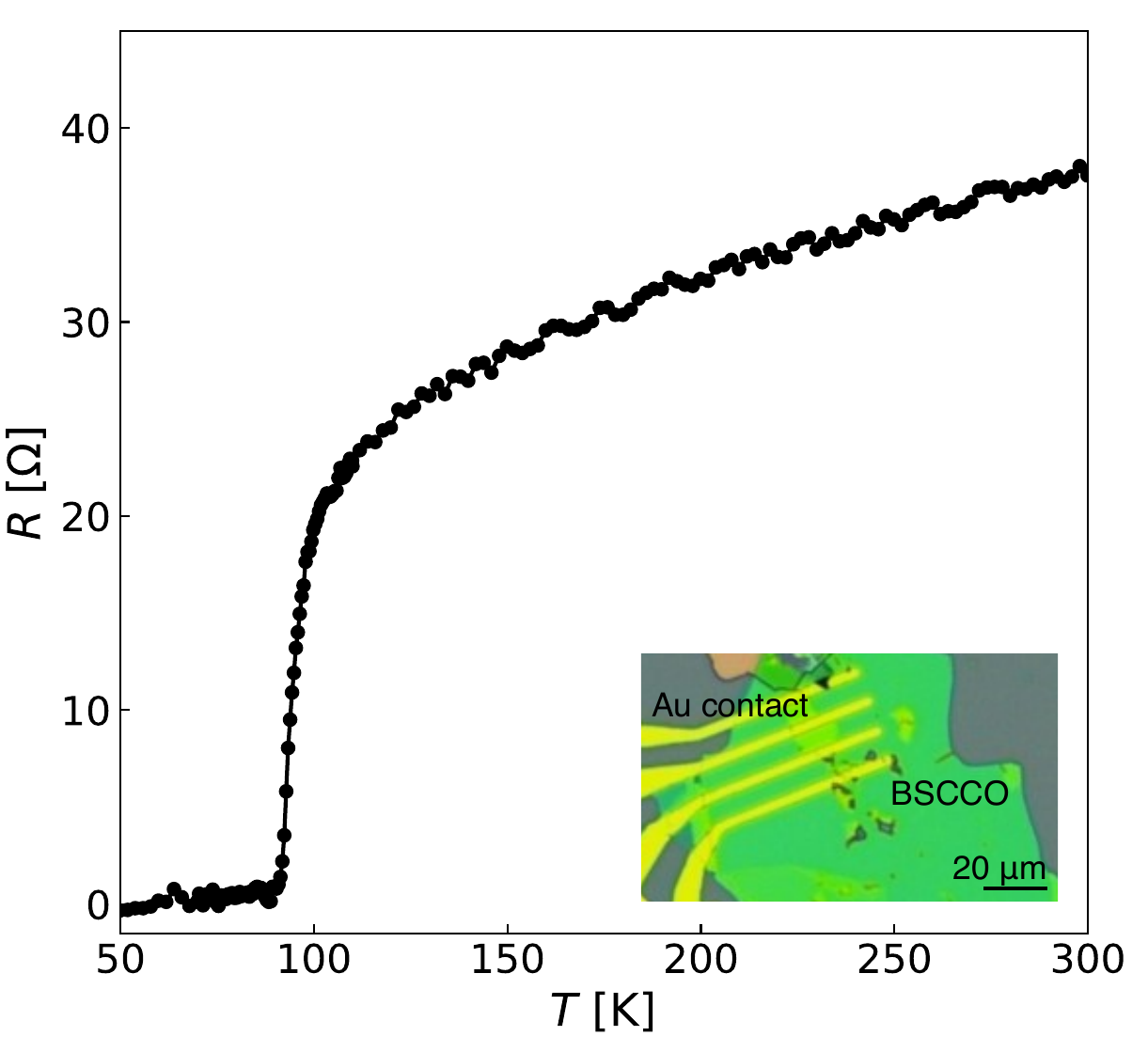}
    \caption{Resistance as a function of temperature of a BSCCO flake that was transferred using PDMS.}
    \label{flake1}     
\end{figure*}

We have placed the BSCCO flakes on top of the microwave resonator made of niobium using the cryogenic transfer technique in the glovebox (see main text). 
Right after the sample was taken out of the glovebox, we have wirebonded the sample to a circuit board. 
Then the device was immediately loaded in a dilution fridge within half an hour. 
To test the quality of the flake, we have performed the same procedure while transferring a 40 nm-thick BSCCO flake onto gold contacts for transport measurement. Figure \ref{flake1} shows a superconducting transition at $\SI{90}{\K}$, comparable to that of a bulk BSCCO crystal. In Ref. \cite{zhao2023time}, BSCCO flakes placed on a contact using the same stacking technique have shown similar results.

\newpage
\section{Measurement Setup}

\begin{figure*}[h]
	\begin{center}
		\includegraphics[scale=0.75]{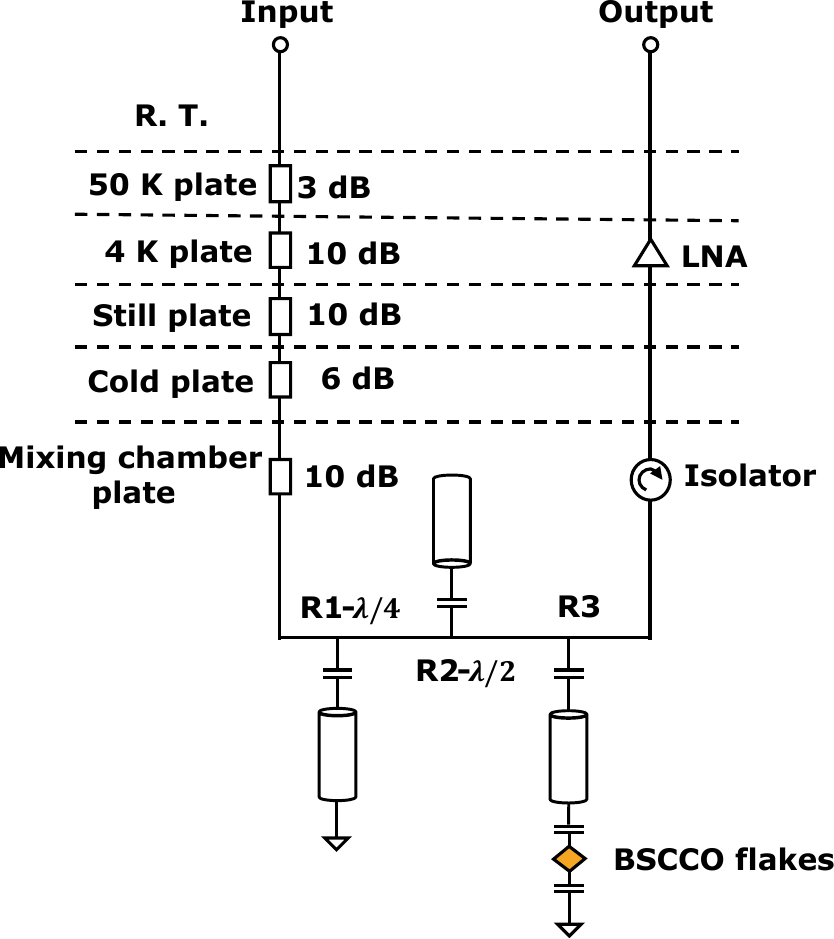}
        \caption{\textbf{Diagram of the microwave characterization setup}. }
		\label{fig7}

	\end{center}
\end{figure*}

All devices were measured in a BlueFors LD400 dilution refrigerator with a base temperature of approximately 30 mK measured on the sample holder. 
A diagram showing the layout for our cryostat and measurement lines is given in Figure \ref{fig7}. 
All measurements were conducted with a P9373A Keysight Streamline Series USB Vector Network Analyzer.

\section{Scanning Transmission Electron Microscopy images}

\begin{figure*}[h]
	\begin{center}
		\includegraphics[width=4.5in]{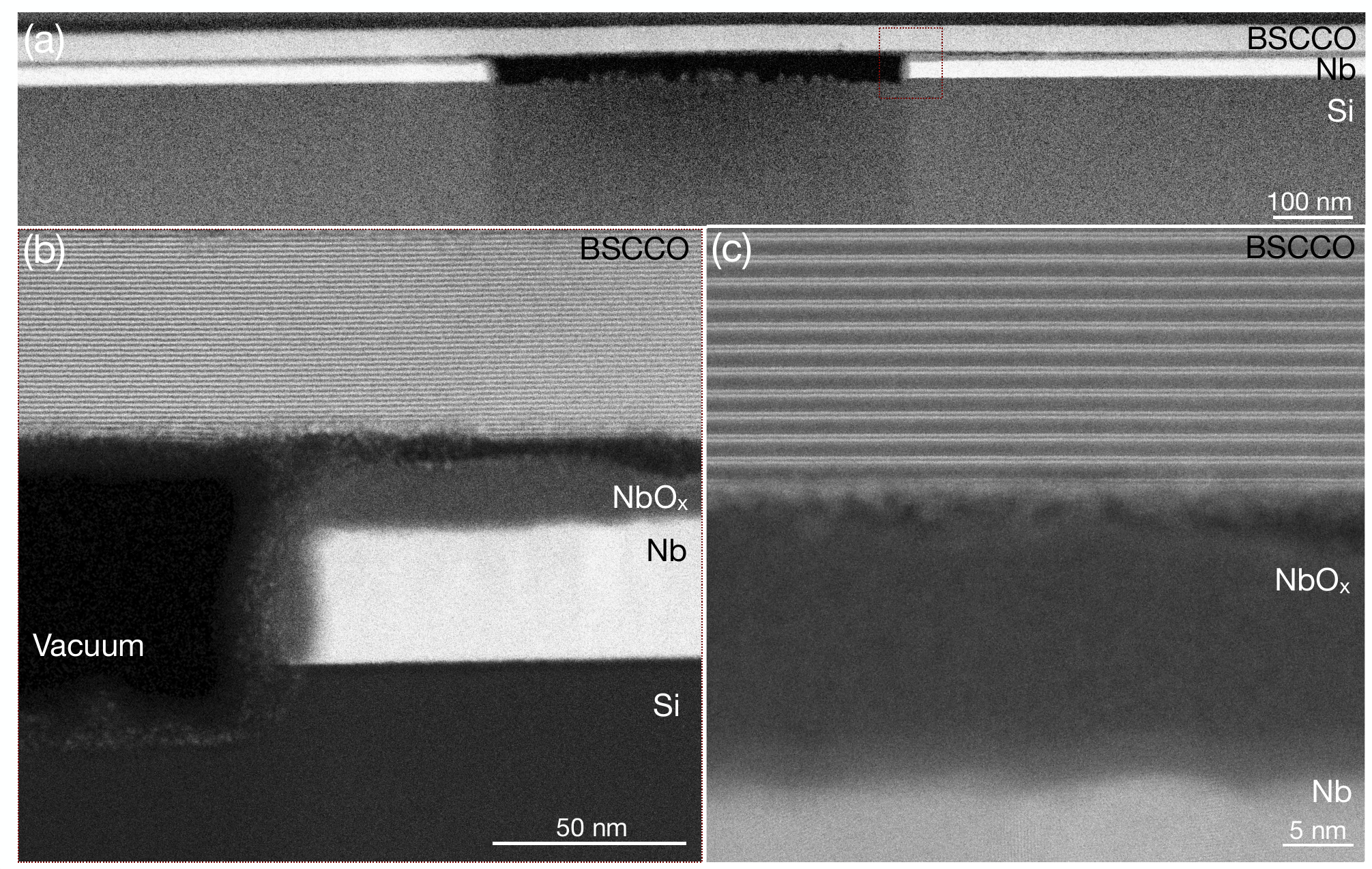}
        \caption{\textbf{TEM image of hybrid device coupled to a BSCCO flake without silicon capping}. (a) STEM image of the overall structure of the BSCCO flake coupled to the niobium resonator without a capping layer on the niobium film. (b) Zoomed-in STEM image at the red dashed box in (a). (c) Close up of the interfaces of BiO atomic layers / niobium oxide layer / niobium.
        Without silicon capping, a rough 15-nm thick niobium oxide layer is formed on the surface of niobium. The BSCCO layers show extensive degradation when in closer contact with the niobium oxide layer.
        }
		\label{fig8}

	\end{center}
\end{figure*}

We analyzed the structural details of the resonator coupled to the BSCCO flake by sectioning the hybrid devices and using  Scanning Transmission Electron Microscopy (STEM) on the region where the BSCCO contacts the resonator.
Figure \ref{fig8}(a) shows a comprehensive STEM image of the resonator coupled to the BSCCO flake, highlighting the overall structure and coupled interface. In the initial devices that we fabricated the BSCCO flake was landed directly on the Niobium resonator.
However, as shown in Figure \ref{fig8}(b) and (c), the niobium oxidation resulted in a $\text{NbO}_{x}$ layer, which, due to surface defects, degraded the coupling with BSCCO.

\begin{figure*}[h]
	\begin{center}
		\includegraphics[width=4.5in]{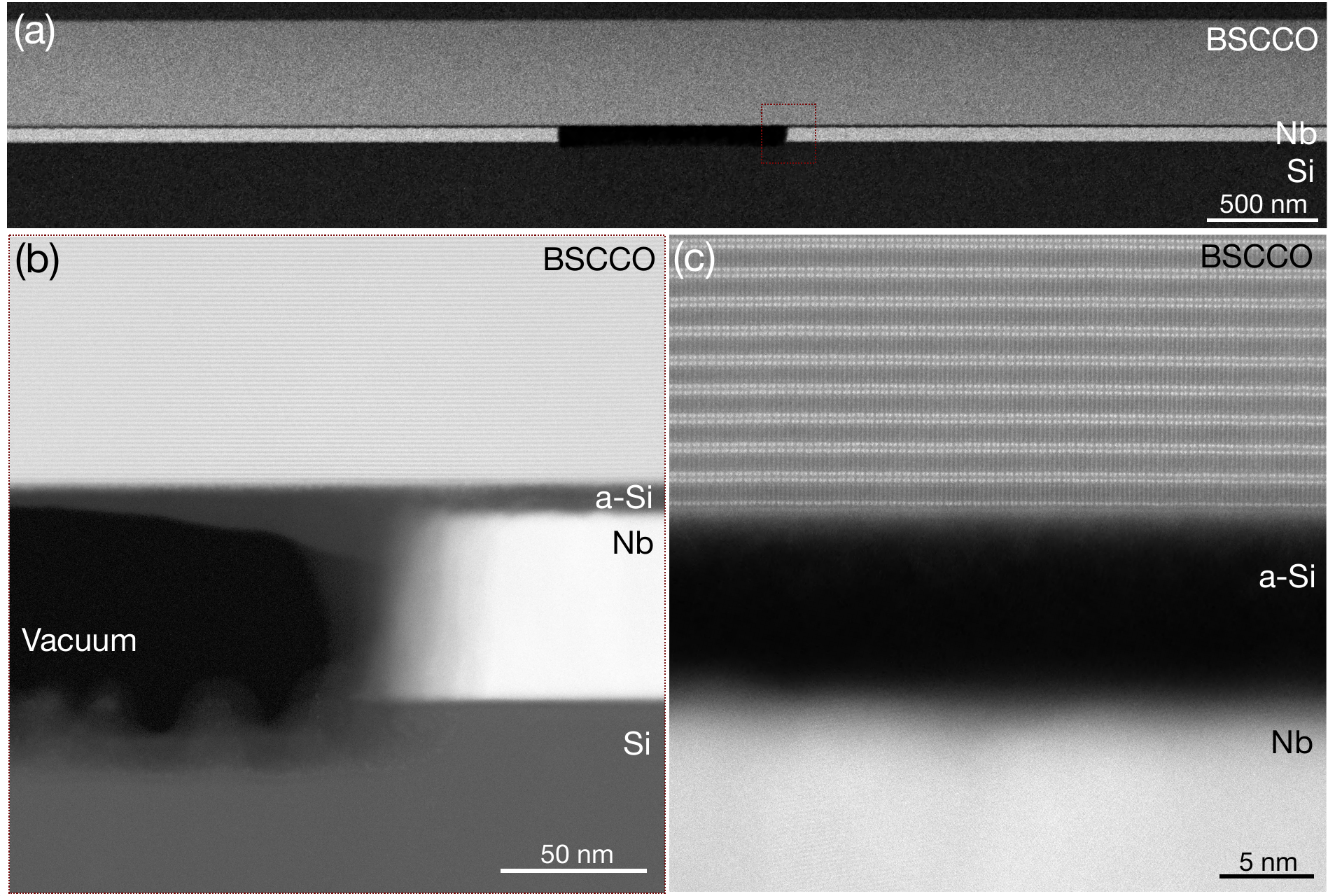}
        \caption{\textbf{TEM of hybrid device coupled to a BSCCO flake with silicon capping}. (a) STEM image of the overall structure of the BSCCO flake coupled to the niobium resonator with silicon capping. (b) Zoom-in at the red dashed box of the cross section. (c) Zoom-in of the interfaces of BiO atomic layers of BSCCO flakes/amorphous silicon/niobium, emphasizing the pristine interface between the flakes and resonator. The niobium oxidation was mitigated with a 6-nm-thick amorphous silicon capping layer.
        }
		\label{fig9}

	\end{center}
\end{figure*}

To address this issue, we deposited a 6-nm-thick amorphous silicon capping layer immediately after the niobium deposition, as shown in Figure \ref{fig9}. 
This modification mitigates the surface defects and enhances the coupling between the niobium resonator and the BSCCO flake, demonstrating an improvement in the device performance.

\newpage

\section{Measuring Internal Quality factor $Q_{int}$}

In this paper, we used a general model to analyze the complex scattering coefficient ($S_{\text{21}}$) of a notch type resonator\cite{gao2008physics,probst2015efficient} :
\begin{align}
S_{21}^{\text {trans}}(f)=\frac{\left(Q_{tot} /\left|Q_{c}\right|\right) e^{i \phi}}{1+2 i Q_{tot}\left(f / f_{r}-1\right)}
\end{align}

where $f$ denotes the measured frequency, $f_{r}$ the resonance frequency, $Q_{tot}$ the total quality factor and $\left|Q_{c}\right|$ the absolute value of the coupling quality factor, and $\phi$ quantifies the impedance mismatch.

\section{Microwave response of different hybrid resonators coupled to BSCCO}

\begin{figure*}[h]
	\begin{center}
		\includegraphics[scale=1.1]{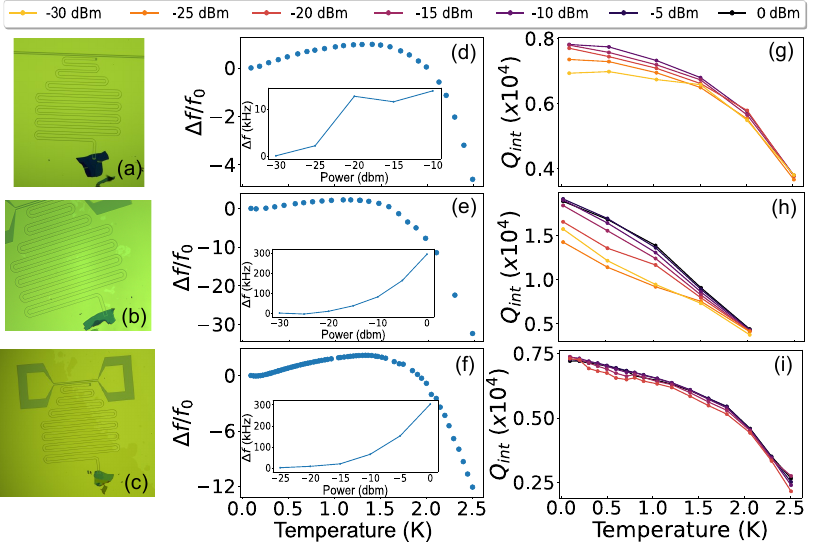}
        \caption{\textbf{Microwave response of hybrid BSCCO devices}. (a-c) Optical micrographs of the three resonators coupled to a BSCCO flake. (d-f) Temperature dependence of the resonance frequencies, respectively to the devices in (a-c). 
        (g-i) Analysis of the internal quality factor at different power levels, further confirming the enhanced performance with the silicon capping layer.}
		\label{fig10}
          
	\end{center}
\end{figure*}

In this supplementary section, we present additional devices of hybrid resonators coupled to BSCCO flakes (Figure \ref{fig10}), focusing on the temperature dependence of their resonance frequencies and their internal quality factors at different microwave power levels. 
For each of the resonators, we observed a shift in the resonance frequency as a function of temperature, which fits well with the hybrid model of quasiparticle excitation and interaction with a TLS reservoir. 
This behavior aligns with the trends in the hybrid BSCCO device (R3) investigated in the main paper, where the BSCCO flake influences the resonator's performance in a similar manner.
We further analyzed the internal quality factor of the resonators at various applied power levels.
The additional measurements show that the frequency upshift as well as the positive nonlinearity are robust across multiple devices.

\begin{table}[h]
\centering
\begin{tabular}{l|r|c|c|c}
\hline
Resonator & Internal Quality Factor & Quality Factor of TLS & Thickness of BSCCO flake (nm)& Si Capped \\
\hline
R3 & $3 \times 10^{4}$ & $2256$& $450$& Yes \\
Supp.A & $9.1 \times 10^{3}$ & $4450$& $68$& No \\
Supp.B & $1.9 \times 10^{4}$ & $1527$& $55$& No \\
Supp.C & $7.6 \times 10^{3}$& $1651$& $83$& No \\
\hline
\end{tabular}
\caption{\label{tab:1} Summary of measured devices. The internal quality factor $Q
_{int}$, the quality factor from the fit to the TLS model $Q_{\mathrm{TLS}}$, and the thickness of the flake are shown for hybrid device R3 from the main text and for the three supplemental devices shown in Figure \ref{fig10}.}
\end{table}

Table \ref{tab:1} gives a summary of the measurements extracted from the different devices. We observe that the internal quality factor is raised with the introduction of a silicon capping layer. However, the TLS quality factor seems unaffected by the change in the internal quality factor and the presence of a silicon capping layer.

\section{Power dependence measurements on the scattering parameters and photon number calibration}

\begin{figure*}[h]
	\begin{center}
		\includegraphics[scale=0.5]{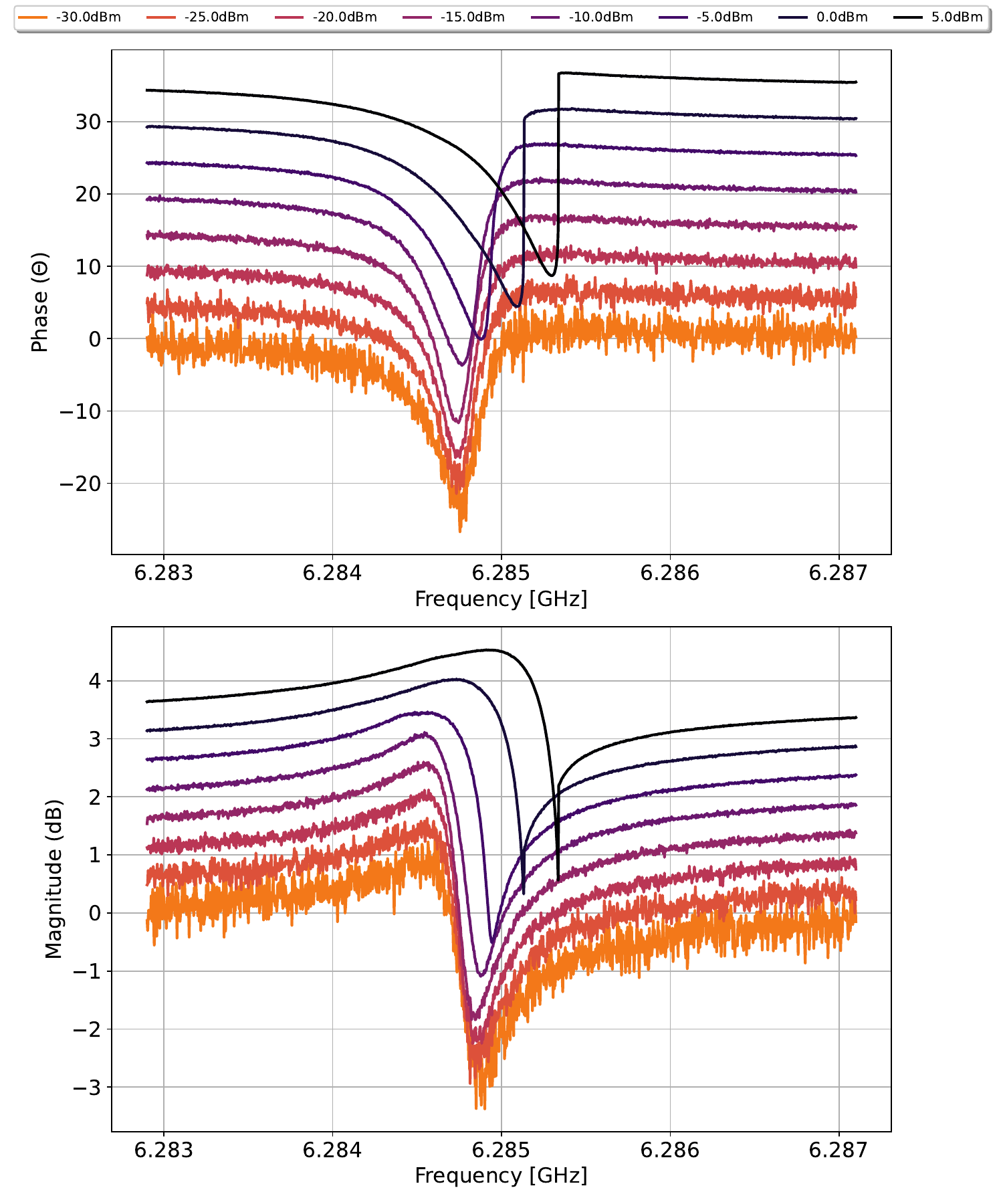}
        \caption{\textbf{Nonlinearity of hybrid resonator R3.}  The phase and magnitude of the transmission spectrum of the BSCCO flake hybrid resonator for varying inpur powers.}
		\label{fig11} 
          
	\end{center}
\end{figure*}

$S_{21}$ measurements of the device R3 at various driving powers are shown in Figure \ref{fig11}. A positive frequency shift with increasing power can be observed, as well as a tilt in the resonance peak. This shift in the resonance shape is observed in the Duffing oscillator and is an indication of nonlinearity and entry into the bifurcation regime.

To quantify the nonlinearity in natural units of $\mathrm{Hz/photon}$, we convert the applied driving power into the number of photons in the resonator by using \cite{clerk2010introduction}:
\begin{align}
\bar{N} & = {P_{in}\frac{4 Q_{total}^2}{\hbar\omega_{r}^2Q_{c}}}
\label{whole_eq2}
\end{align} 
where $\bar{N}$ is the average photon number inside the resonator, $P_{in}$ is the input power at the transmission line port, $Q_{total}$ is the total quality factor of the resonator, $Q_{c}$ is the coupling quality factor between the  transmission line and resonator and $\omega_{r}$ is the resonance frequency. 
The largest uncertainty in the photon number estimate comes from the difficulty in accounting for the total attenuation between the generator and the transmission line and the value of $P_{in}$. 
We estimate a total attenuation 80 dB, but the uncertainty in this value indicates that our calibrated photon number should be considered up to one order of magnitude.

\section{TLS effects in the bare niobium resonator}

As mentioned in the main text, the bare niobium resonator R2 also shows a slight frequency upshift and its quality factor increases significantly with drive power. Here, we use a combined quasiparticle and TLS model to quantify this effect.

\begin{figure*}[h]
	\begin{center}
		\includegraphics[scale=0.7]{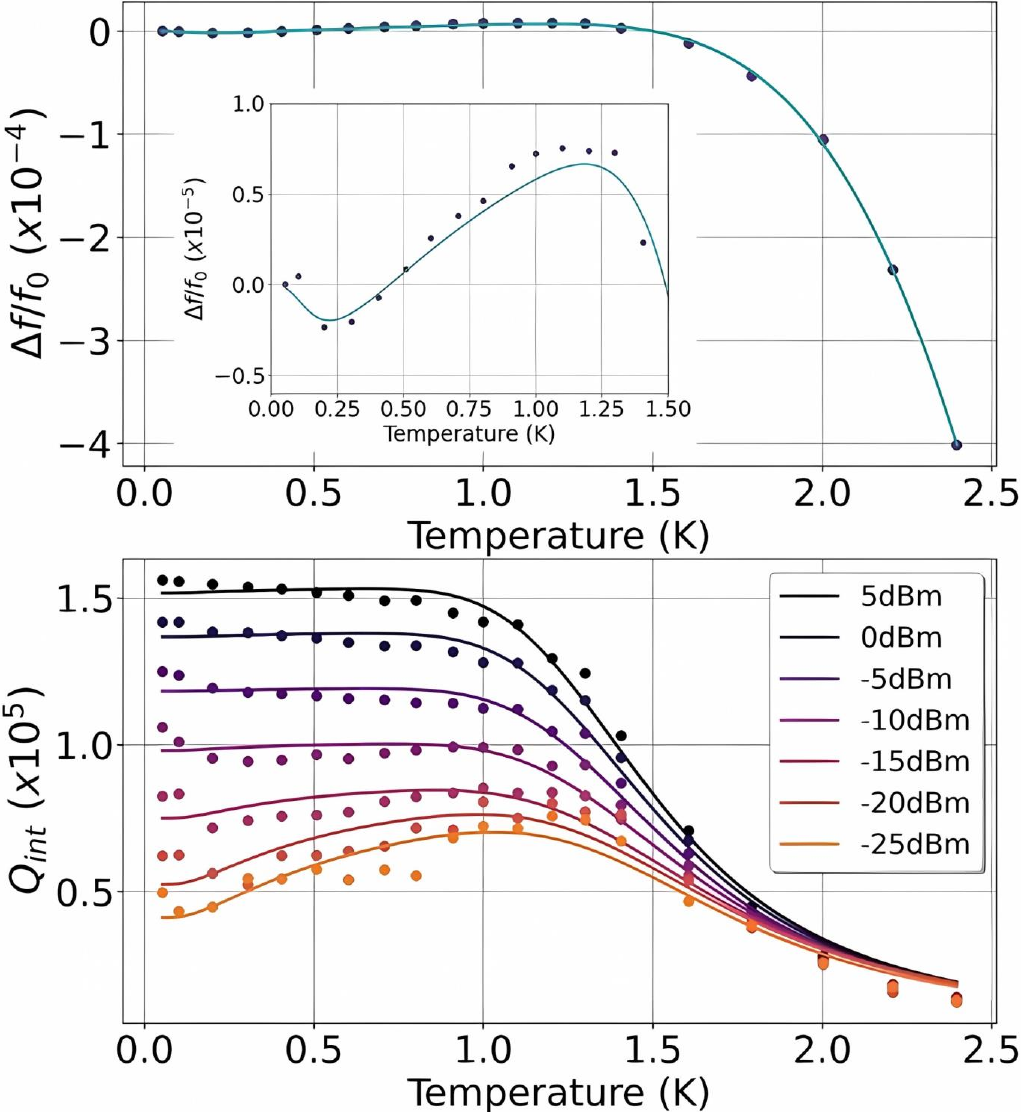}
        \caption{\textbf{Parametrizing TLS effects in the bare niobium resonator.} The top panel shows the resonance frequency of resonator R2 vs. temperature, similar to main text Figure 3 but here fit with a combined quasiparticle-TLS model. The inset shows a zoom-in focusing on the frequency upshift at low temperatures. The bottom panel shows the internal quality factor of resonator R2 vs. temperature at different drive powers, fit with a combined quasiparticle-TLS model explained in the text.}
		\label{fig12} 

	\end{center}
\end{figure*}

In this model, the resonance frequency shift is given by:
\begin{align}
\frac{\delta f(T)}{f_{0}} &= \left(\frac{\delta f(T)}{f_{0}}\right)_{\mathrm{TLS}} + \left(\frac{\delta f(T)}{f_{0}}\right)_{\mathrm{QP}}
\end{align}
where
\begin{align}
\begin{split}
\left(\frac{\delta f(T)}{f_{0}}\right)_{\mathrm{TLS}} =\frac{1}{\pi Q_{\mathrm{TLS}}} \operatorname{Re}\biggl[\Psi\left(\frac{1}{2}+i \frac{\hbar \omega}{2 \pi k_{B} T}\right) 
-\ln \left(\frac{\hbar \omega}{2 \pi k_{B} T}\right)\biggr] 
\end{split}
\end{align}
is the TLS bath contribution to the frequency shift as discussed in the main text \cite{gao2008physics}.
The frequency shift caused by quasiparticles is given by \cite{prozorov2006magnetic}:
\begin{align}
\left(\frac{\delta f(T)}{f_{0}}\right)_{\mathrm{QP}}=A \sqrt{\frac{2 \pi \Delta_{0}}{T}} \mathrm{e}^{-\Delta_{0} / T}
\end{align}
where $A$ is a constant accounting for the participation of the kinetic inductance, and $\Delta_{0}$ is the superconducting gap.
As shown in Figure \ref{fig12}, by fitting the bare niobium resonator with the combined model, we can quantify $Q_{\mathrm{TLS}}$  as $3.2\times10^{4}$, which is comparable with the internal quality factor of the bare resonator at low drive powers.

Furthermore, we use a combined model to examine the temperature-dependent internal quality factor at different drive powers:
\begin{align}
\frac{1}{Q_{\mathrm{int}}}=\frac{1}{Q_{\mathrm{TLS}}(T)}+\frac{1}{Q_{\mathrm{QP}}(T)}+\frac{1}{Q_{\text {other }}}
\end{align}
where the $Q_{\mathrm{TLS}}(T)$ is the TLS contribution the quality factor, $Q_{\mathrm{QP}}(T)$ is the quality factor dominated by quasiparticle losses, and $Q_{\text {other }}$ accounts for additional temperature-independent losses. 
The temperature dependence of the quality factors is given by\cite{mattis1958theory,gao2008physics}:
\begin{align}
Q_{\mathrm{TLS}}(T)=Q_{\mathrm{TLS}}/\tanh \left(\frac{\hbar \omega}{2 k_{B} T}\right)
\end{align}
and
\begin{align}
Q_{\mathrm{QP}}(T)=Q_{\mathrm{QP},0} \frac{e^{\Delta_{0} / k_{B} T}}{\sinh \left(\frac{\hbar \omega}{2 k_{B} T}\right) K_{0}\left(\frac{\hbar \omega}{2 k_{B} T}\right)}
\end{align}
where $K_{0}(x)$ is the zeroth order modified Bessel function of the second kind.

Figure \ref{fig12} shows the fit of the bare niobium resonator quality factor to the combined model at various temperatures. The fitted parameter $Q_\mathrm{TLS}$ increases consistently with the drive power, from $7.2\times10^{4}$ at $-25\ \mathrm{dBm}$ to effectively infinity at the highest driving power, where quasiparticle loss alone can account for the quality factor temperature dependence. 

 $Q_\mathrm{TLS}$ extracted from the quality factor measurement is consistent with the one extracted from the resonance frequency. 
This agreement shows that a broad frequency TLS reservoir accounts for the behavior of the bare niobium resonator, as has been observed in previous resonator studies and attributed to dielectric losses. 
We note a significant difference between the bare resonator and the hybrid resonator, where the TLS bath shows a large frequency response with minimal effect on the quality factor, possibly due to mostly a reservoir of primarily off-resonant TLS defects. Therefore, it is evident that the source of the anomalous upshift and nonlinearity in the hybrid device is separate from typical TLS effects in superconducting resonators an due to a new mechanism within the BSCCO flake itself.

\section{Distinguishing the effect of residual PDMS}

\begin{figure*}[h]
	\begin{center}
		\includegraphics[scale=0.26]{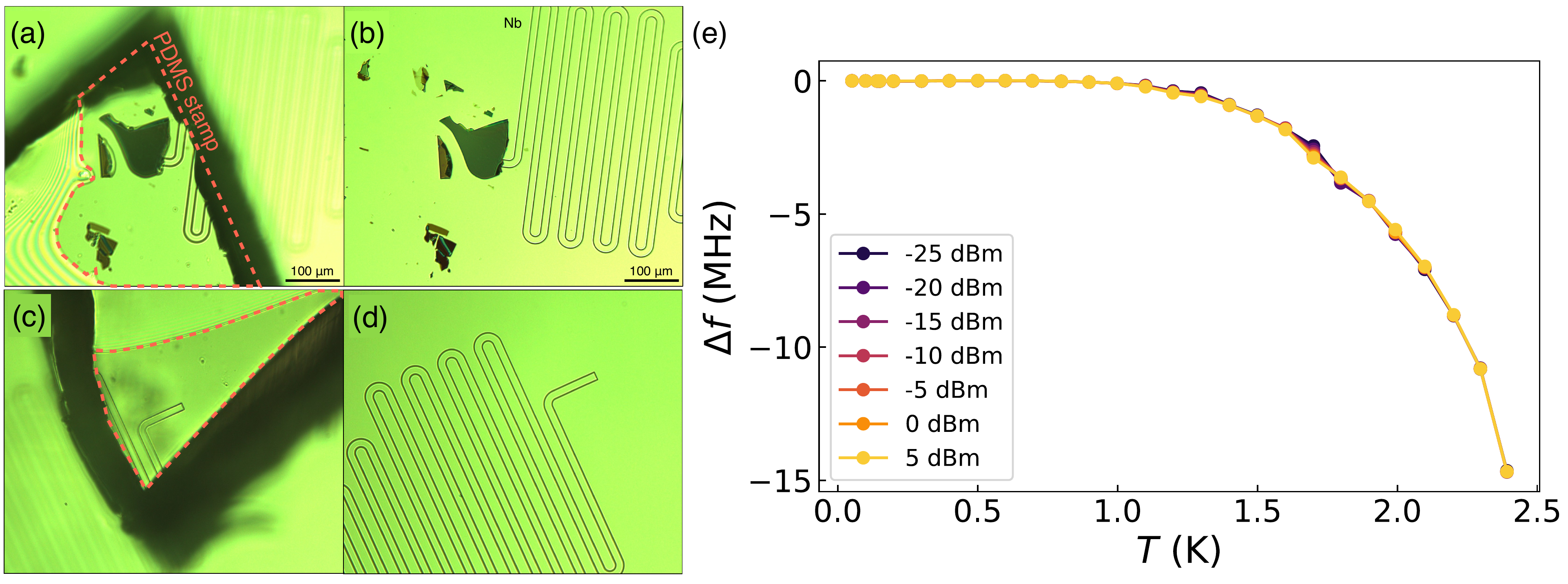}
        \caption{\textbf{Control experiment of PDMS}  (a-b) Optical micrographs of the transfer process using the PDMS stamp for a flake investigated in the manuscript and the area marked as a red line was in contact with the stamp on the niobium resonator. (c) Optical micrographs of the PDMS stamp attached on the niobium resonator and (d) the niobium resonator after the stamp was detached. (e) The temperature dependent resonance frequency at different applied powers.}
		\label{fig13}

	\end{center}
\end{figure*}

A potential concern of our measurements is that the TLS defects possibly originate from residual PDMS left during the flake transfer procedure. To test this, we have put a PDMS stamp on a niobium resonator without any flake, using the same transfer procedure (see Figure \ref{fig13}a-d). Then we have detached the PDMS stamp and measured the microwave resonance frequency as a function of temperature. 

As can be seen in Figure \ref{fig13}e, the resonance frequency gradually drops when the temperature is increased, and appears power independent. This behavior is consistent with a bare resonator and does not show the strong TLS signal we have observed in the hybrid resonators.
\newpage
{}